\documentclass{aa}

\usepackage{graphicx}
\usepackage{txfonts}
\usepackage{lipsum}
\usepackage{subcaption}         
                                
\usepackage{lscape}             
                                
\usepackage{placeins}
\usepackage{amsmath}	
\usepackage{multirow}
\usepackage{booktabs}
\usepackage{adjustbox}
\usepackage{hyperref}
\usepackage{float}

\begin{document}

   \title{Probabilistic Stellar Mass and Radius Estimation with Bayesian Hierarchical Stacking}

   \subtitle{}

   \author{Maxence Lamirel\inst{1,2}\corrauth{maxence.lamirel.research@gmail.com}       
        \and Andr\'es Moya\inst{2,3}
        \and Roberto J. L\'opez-Sastre\inst{4}
        \and Luis Manuel Sarro\inst{5}}

   \institute{Centro de Astrobiología, CSIC-INTA, Camino Bajo del Castillo s/n, 28692 Villanueva de la Cañada, Madrid, Spain
   \and Dpt. Astronomia i Astrof\'isica, Universitat de Val\`encia, Av. Vicent Andr\'es Estell\'es, 19, 46120 Burjassot, Spain
   \and Observatori Astron\`omic, Universitat de Val\`encia, C/ Cat. Jos\'e Beltr\'an 2, 46980 Paterna, Spain
   \and GRAM Research Group, Department of Signal Theory and Communications, University of Alcal\'a, 28805 Alcal\'a de Henares, Spain
   \and Departamento de Inteligencia Artificial, ETSI Inform\'atica, UNED, Juan del Rosal, E-16, 28040 Madrid, Spain}

   \date{Received June 9, 2026}
 
  \abstract 
   {Accurate stellar parameters are essential for the characterisation of exoplanets, yet most host stars lack direct measurements of mass and radius. These parameters set the inferred planetary properties, including radius, mass, and bulk density, with important implications for exoplanet physics, including potential habitability and the search for biomarkers.}
   {We aim to predict stellar masses and radii from commonly available observables and to quantify uncertainties and systematic effects in a way that is consistent with stellar observations. By not relying on a single Bayesian machine learning model, we aim to obtain more robust predictive distributions and reduce sensitivity to model misspecification and outliers.} 
   {We develop a Bayesian machine-learning framework that combines three probabilistic models through Bayesian hierarchical stacking (BHS): a Bayesian additive regression tree model, a two-process heteroscedastic sparse Gaussian process, and a hierarchical Bayesian neural network. The method propagates measurement uncertainties, accounts for correlations among inputs, and aggregates model variance into final predictive distributions.}
   {On a held-out test set the method attains a mean absolute relative difference (MARD) of 3.5\% for mass and 2.1\% for radius, improving on empirical-relation-based methods and standard machine-learning baselines. For 44 Kepler LEGACY stars, the BHS masses differ from estimates obtained with the BAyesian STellar Algorithm (BASTA) by a MARD of 4.8\%, compared with 8.5\% between the empirical relations and BASTA. Applied to about 2400 exoplanet host stars, the predictions differ from literature values by a MARD of 2.7\% in both mass and radius. Residual analysis reveals mild underestimation for cool dwarfs, consistent with sparse training coverage, and positive offsets near the main-sequence turn-off that may reflect priors in isochrone-based catalogues. Internal consistency with $g=GM/R^2$ is greater for the BHS predictions, with a scatter in $\Delta\log g$ lower by a factor of three than that of the literature values.}
   {}


   \maketitle
   \nolinenumbers

\section{Introduction}

The characterisation of exoplanetary systems is fundamentally limited by our knowledge of their host stars. The two primary detection techniques—radial velocity and transit photometry—yield planetary masses and radii, respectively, and both depend intrinsically on stellar parameters. Accurate determination of stellar properties is therefore essential for deriving reliable planetary characteristics and for interpreting population-level trends, such as the small planet radius gap \citep{Petigura_2022}.

However, direct measurements of stellar mass and radius are available only for a limited subset of stars, mainly via asteroseismology \citep{asteroseis}, eclipsing binaries \citep{torres_2010}, and interferometric observations \citep{interfero} in the case of the radius, while the majority of stars require indirect estimation methods. For example, empirical relations between easy-to-measure parameters such as $T_{\text{eff}}$, $\log g$, [Fe/H] and mass or radius have been devised using eclipsing-binary data \citep{torres_2010}. These relations are used to study exoplanet host stars \citep{santos_2016}. \citet{moya1} presented 38 empirical relations between $T_{\text{eff}}$, $\log g$, [Fe/H], $\rho$, $L$ to be used as independent variables for the same purpose. This was done with an improved data sample of 700 main-sequence stars that included eclipsing binaries, asteroseismic and interferometric data.

Grid-based modelling remains the standard approach for estimating stellar masses and radii in exoplanet studies. These methods compare observed stellar properties with theoretical isochrones derived from stellar structure and evolution models, often using Bayesian techniques to identify the range of models compatible with the observations \citep[e.g.][]{bazot_2012, isochrone, Petigura_2022}. Modern implementations include the BAyesian STellar Algorithm (BASTA), which can combine spectroscopic, photometric, astrometric, and asteroseismic constraints, and Asteroseismic Inference on a Massive Scale (AIMS), which is primarily designed for inference using asteroseismic frequencies together with classical constraints \citep{basta,aims}. Despite their success, grid-based models remain limited by uncertainties in key stellar physics, such as convection, overshooting, nuclear reaction rates, and the effects of rotation and magnetic fields. Stellar radii can also be inferred more directly from Gaia parallaxes and bolometric luminosities via the Stefan–Boltzmann relation \citep{cksVII}, which generally yields higher precision than spectroscopic grid fitting. Incorporating Gaia constraints as priors further improves the robustness of isochrone-based inferences \citep{berger2020gaiakeplerstellarpropertiescatalog}.

Machine learning has also been increasingly applied to stellar characterisation, both in the analysis of asteroseismic data \citep{Verma_2016} and within grid-based modelling frameworks \citep{maistep}. \citet{moya2} predicted stellar masses and radii using several machine-learning algorithms and showed that combining them through stacking yielded closer agreement with held-out values. However, such methods generally lack a rigorous treatment of uncertainties, limiting their reliability for detailed astrophysical inference.

Building on the work of \citet{moya2} and leveraging the stellar database described in \citet{moya1}, we propose a Bayesian machine-learning framework for estimating stellar masses and radii from spectroscopic and photometric parameters ($T_{\text{eff}}$, $\log g$, [Fe/H], and $L$). Our approach employs probabilistic programming with next-generation Markov chain Monte Carlo (MCMC) algorithms to define three Bayesian models: an errors-in-variables Bayesian additive regression trees (BART) model, a two-process heteroscedastic sparse Gaussian process (TP-HSGP), and a hierarchical Bayesian neural network (HBNN). These models are subsequently combined through Bayesian hierarchical stacking, allowing us to account for correlations among input parameters, propagate measurement uncertainties, and incorporate model variance into the final estimates.

Our framework differs from grid-based inference in that, once trained, it does not require the selection or interpolation of stellar-evolution tracks, nor the specification of stellar input physics or evolutionary priors for each target. Instead, it learns the relation between commonly available observables and the high-quality benchmark determinations in the curated training sample described in Sect.~\ref{sec:data}. These reference values are derived from asteroseismology, eclipsing binaries, and interferometry. Although well constrained, they are not treated as true or fully model-independent values, and their method-dependent systematics may propagate into our predictions. The model provides probabilistic mass and radius estimates from a small number of commonly available parameters, including for stars without detected oscillations. We regard this as complementary to grid-based inference, particularly when detailed asteroseismic constraints are unavailable. 

Our work evaluates the framework in three complementary settings. We first assess its predictive performance and uncertainty calibration on a held-out test set. We then compare our mass estimates with BASTA results for 44 Kepler LEGACY stars. Finally, we apply the framework to approximately 2400 exoplanet hosts from the NASA Exoplanet Archive \citep{christiansen2025nasaexoplanetarchiveexoplanet}. Comparisons with empirical relations and isochrone-based catalogues allow us to examine systematic differences across the Hertzsprung--Russell diagram and the internal consistency of the predicted masses and radii through $g=GM/R^2$. Together, these analyses assess the applicability of probabilistic machine learning to the catalogue-scale characterisation of exoplanet host stars.

\section{Data}
\label{sec:data}

\subsection{Model-development sample}

\begin{figure}
    \centering
   \includegraphics[width=\columnwidth]{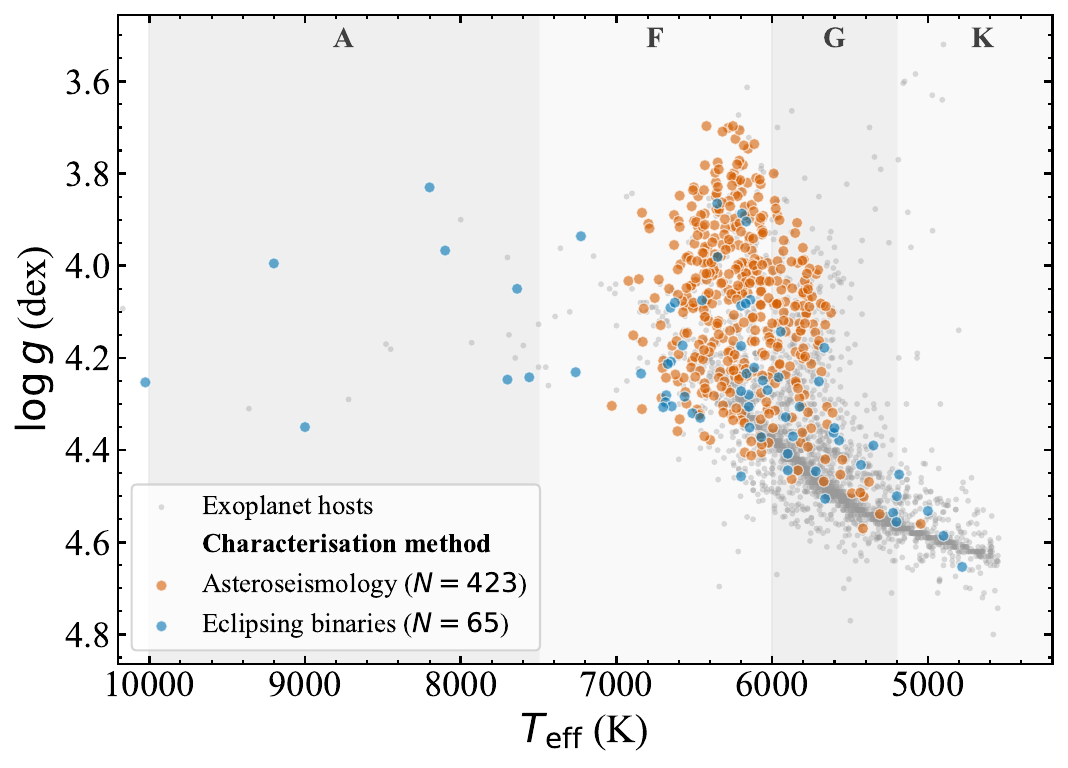} 
   \caption{Kiel diagram showing the exoplanet-host sample selected from the NASA Exoplanet Archive (grey) and our 488-star data sample from \citet{moya1}, divided according to the principal characterisation methods. The shaded regions indicate approximate spectral-type ranges based on $T_{\rm eff}$.}
   \label{fig:HR}
\end{figure}

The model-development sample used in this work is drawn from the compilation of \citet{moya1}, which initially contained approximately 2000 entries collected from multiple studies. The largest contributions are the 222 stars from \citet{eker} and the 397 stars from \citet{serenelli}. The compilation pairs commonly available stellar observables with published mass and radius determinations while avoiding, where possible, quantities derived from the same parameters.

The reference masses and radii are obtained primarily through asteroseismic modelling and analyses of detached eclipsing binaries. Interferometric observations provide additional independent constraints on stellar radii. These methods yield well-constrained reference determinations but remain subject to method-dependent assumptions and systematic uncertainties.

The values of $[\mathrm{Fe/H}]$ and $T_{\mathrm{eff}}$ are compiled primarily from spectroscopic observations, with some $T_{\mathrm{eff}}$ determinations obtained photometrically. Luminosities are either taken from the original studies or derived using the Virtual Observatory (VO) Spectral Energy Distribution Analyser \citep{bayo}, based on bolometric fluxes and distances \citep{gaia}. The reported $\log g$ values are compiled from the original sources and, in most cases, are not calculated directly from the paired mass and radius values.

Duplicate entries were filtered according to the precision of the published determinations, leaving 726 main-sequence stars, predominantly of spectral type G \citep{moya1}. We retain the 488 stars with measurements and reported uncertainties for all of $M$, $R$, $T_{\mathrm{eff}}$, $L$, $[\mathrm{Fe/H}]$, and $\log g$. Further details of the source catalogues and data-quality criteria are provided by \citet{moya1}.

Fig.~\ref{fig:HR} shows that the reference sample covers spectral types A--K, although 97 per cent of the stars are of type F or G. Approximately 87 per cent are characterised through asteroseismology and are predominantly Sun-like pulsators. The eclipsing-binary subsample is distributed more broadly along the main sequence. 

\begin{figure}
    \centering
   \includegraphics[width=\columnwidth]{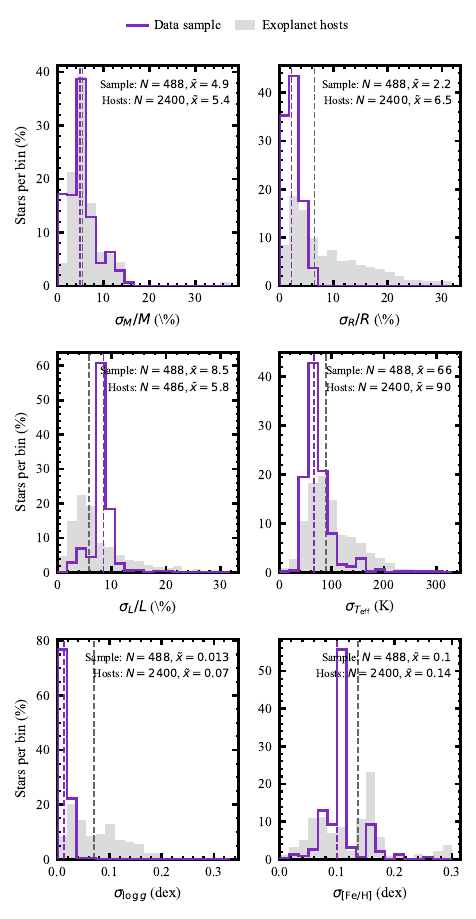} 
   \caption{Distributions of the reported measurement uncertainties for our 488-star sample (purple outlines) and the exoplanet-host comparison sample (grey filled histograms). Uncertainties in mass, radius, and luminosity are expressed as percentages of the corresponding parameter values, whereas uncertainties in $T_{\mathrm{eff}}$, $\log g$, and $[\mathrm{Fe/H}]$ are shown in their reported units. Each histogram is normalised separately, and the dashed lines indicate the corresponding medians.}
   \label{fig:uncert}
\end{figure}

Fig.~\ref{fig:uncert} presents the distributions of the reported uncertainties used in our analysis. In our sample, the median uncertainties are 4.9 per cent in mass, 2.2 per cent in radius, 8.5 per cent in luminosity, 66 K in $T_{\mathrm{eff}}$, 0.013 dex in $\log g$, and 0.10 dex in $[\mathrm{Fe/H}]$. The distributions nevertheless contain extended tails, particularly for some of the literature measurements in the exoplanet-host sample. Because these reported uncertainties are propagated through the Bayesian models, their full distributions, rather than only their typical values, are relevant to the resulting predictive uncertainties.

\subsection{Data split}
\label{sec:split}
The model-development sample is randomly divided into training and test subsets containing 80 and 20 per cent of the stars, respectively. The test subset is held out during model fitting and is used to evaluate predictive performance and uncertainty calibration.

\begin{figure}
    \centering
   \includegraphics[width=\columnwidth]{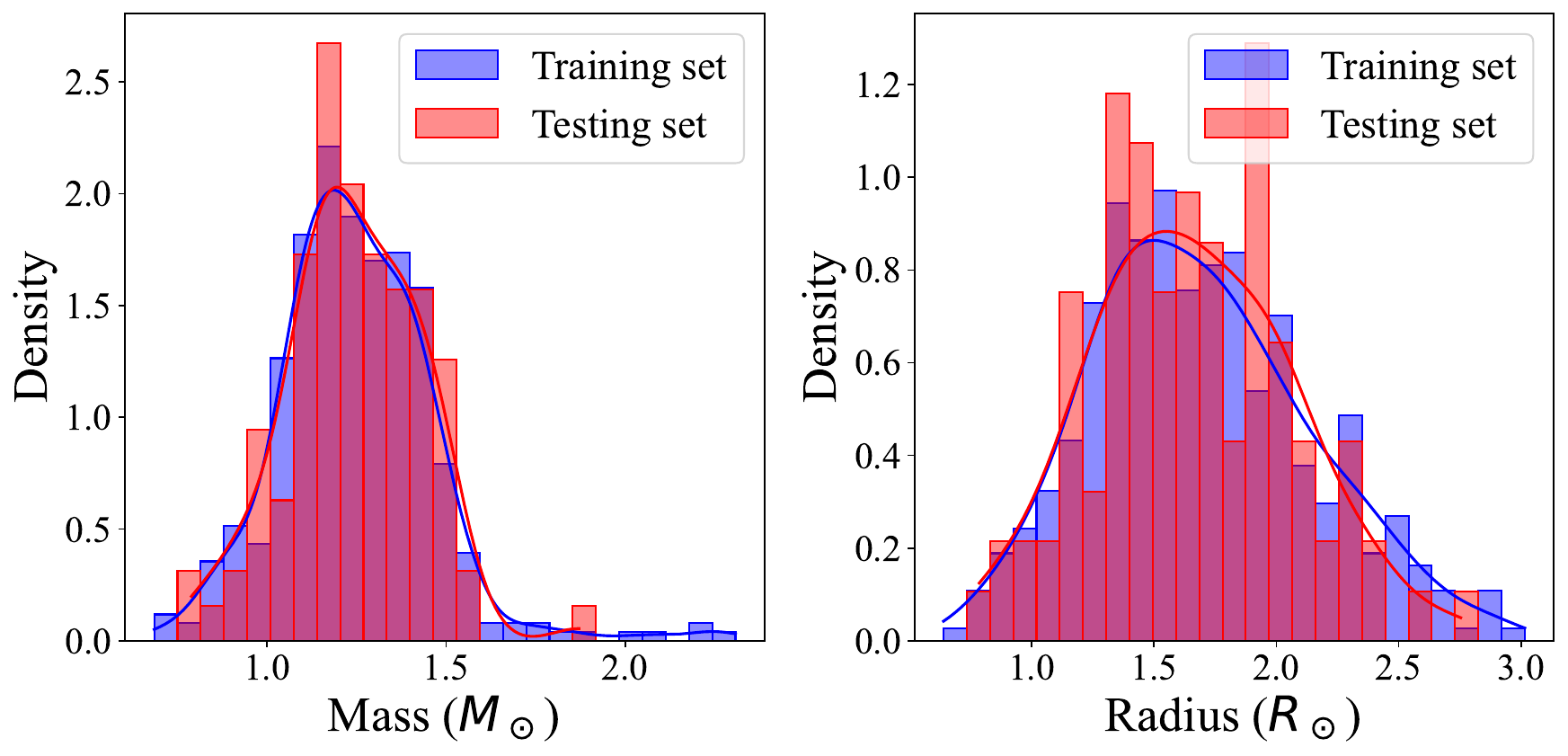} 
   \caption{Histogram of the mass and radius for the training set (blue) and test set (red) alongside the general density of values. The test set gives a good representation of the values in the general dataset.}
   \label{fig:traintest}
\end{figure}

Fig. \ref{fig:traintest} shows the distribution of the mass and radii values respectively for both the testing and training samples. We can see from the density that both sets are statistically coherent and represent all types of values from our dataset. The testing values cover the training set population correctly. 

Before fitting the models, each parameter is standardised to zero mean and unit variance using the mean and standard deviation of the training subset. The reported uncertainties are divided by the corresponding training-set standard deviations. The same transformation is applied to the test set and all external comparison samples. This places the predictors on comparable scales.

\subsection{Exoplanet-host comparison sample}
\label{sec:host_data}
In section \ref{sec:host_pred}, we use our trained models to make mass and radius predictions of host stars using data from the NASA Exoplanet Archive \cite{christiansen2025nasaexoplanetarchiveexoplanet}. We use the default parameter set recommended by the NASA archive for each star. We then retain only stars with values for at least $T_{\text{eff}}$, [Fe/H], $\log (g)$ as well as an estimation of mass and radius. We further limit the dataset to inputs within 5 per cent of our corresponding maximum and minimum training values. These data were downloaded from the NASA archive in August of 2025. 

\begin{table}
    \centering
    \caption{Ranges of literature parameters used in host star predictions and comparison.}
    \begin{tabular}{lcc}
        \toprule
        Parameter & Minimum & Maximum \\
        \midrule
        $M$ ($M_\odot$) & 0.478 & 2.520 \\
        $R$ ($R_\odot$) & 0.613 & 3.210 \\
        $T_{\mathrm{eff}}$ (K) & 4546 & 10170 \\
        $\log g$ (dex) & 3.52 & 4.80 \\
        $[\mathrm{Fe}/\mathrm{H}]$ (dex) & -0.66 & 0.49 \\
        Luminosity ($L_{\odot}$) & 0.16 & 53.00 \\
        \bottomrule
    \end{tabular}
    \tablefoot{Literature values were retrieved from the NASA Exoplanet Archive \citep{christiansen2025nasaexoplanetarchiveexoplanet}.}
    \label{tab:valid_range}
\end{table}

We make predictions for 2400 hosts. The three most represented catalogues are \citet{morton_2016, Q1_Q7_DR25_KOI, macdougall_2023} with 1078, 288 and 53 stars respectively. The ranges of the parameters used as part of our predictions are given in Table \ref{tab:valid_range} and the distributions of these stars across the Kiel diagram in Fig. \ref{fig:HR}. Of these stars, 486 have luminosity values. 

In the majority of cases, stellar masses and radii are obtained through isochrone fitting. This is true for the three most represented catalogues in our dataset, which together account for more than half of the stars. Our comparison therefore mainly evaluates our Bayesian machine-learning framework against modern isochrone-based stellar characterisation methods. In terms of input uncertainties, the exoplanet-host parameters tend to have larger errors than our own data sample apart from luminosity as we can see in Fig. \ref{fig:uncert}.

\section{Machine-learning models}

We estimate stellar masses and radii using an ensemble based on Bayesian hierarchical stacking (BHS; \citealt{wolpert_1992,bhs}). Stacking combines the predictions of several independently trained models, allowing the ensemble to exploit their complementary strengths. In BHS, the contribution of each model can vary across the stellar-parameter space, while uncertainty in both the individual predictions and their combination is propagated to the final estimate.

\begin{figure}
    \centering
    \includegraphics[width=\columnwidth]{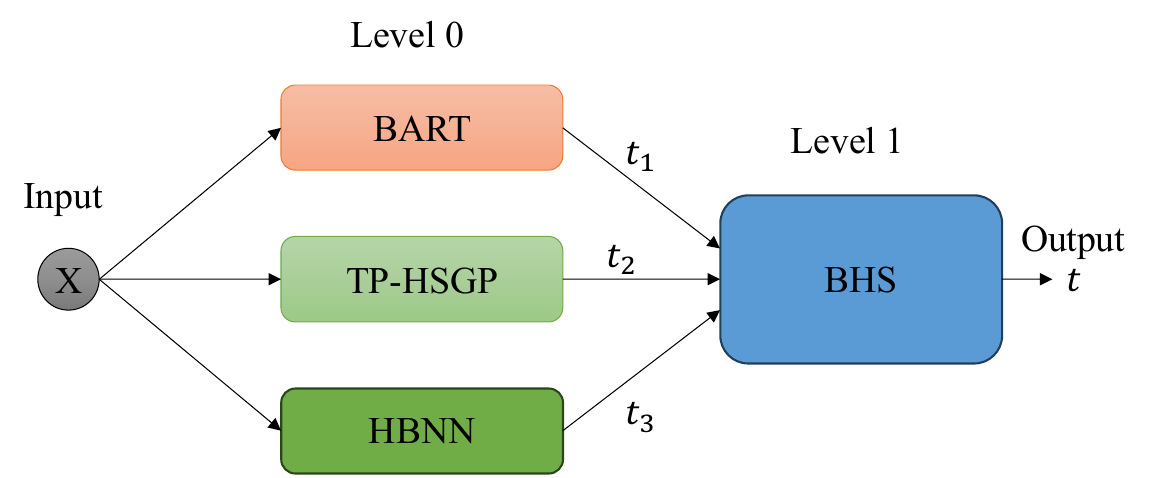} 
    \caption{Architecture of the stacked model. The level-0 ensemble contains three probabilistic predictors: a Bayesian additive regression trees (BART) model, a two-process heteroscedastic sparse Gaussian process (TP-HSGP), and a hierarchical Bayesian neural network (HBNN). Their predictive distributions are combined at level-1 using Bayesian hierarchical stacking (BHS).}
    \label{fig:bhs}
\end{figure}

Our architecture is shown in Fig.~\ref{fig:bhs}. At level-0, three probabilistic models independently predict $M_\star$ or $R_\star$ from the stellar parameters $[T_{\mathrm{eff}}, \log g, [\mathrm{Fe/H}], L]$: Bayesian additive regression trees (BART), a two-process heteroscedastic sparse Gaussian process (TP-HSGP), and a hierarchical Bayesian neural network (HBNN). At level-1, BHS combines their predictive distributions to obtain the final mass or radius estimate.

The three level-0 models were chosen for their complementary statistical properties rather than to represent different stellar-physics prescriptions. BART captures non-linearities and interactions without assuming a globally smooth relation, while the TP-HSGP models a smooth relation between the stellar parameters and the target and allows the noise variance to depend on the reported measurement uncertainties. The HBNN provides a flexible non-linear mapping while explicitly modelling correlations among the latent stellar parameters. All three models provide posterior predictive distributions, but incorporate observational uncertainties through different statistical constructions.

\subsection{Bayesian hierarchical stacking}
\label{sec:bhs}

Bayesian hierarchical stacking generalises standard stacked model averaging by allowing the relative importance of the individual models to depend on position in the predictor space \citep{bhs}. This is useful here because the relative performance of the level-0 models may vary across different regions of stellar-parameter space.

Each level-0 model provides a posterior predictive distribution, $p_k(y\mid\mathbf{x})$, where $y$ denotes either stellar mass or radius. The stacked predictive distribution is

\begin{equation} 
p_{\mathrm{stack}}(y\mid\mathbf{x}) = \sum_{k=1}^{K} w_k(\mathbf{x})\,p_k(y\mid\mathbf{x}),
\label{eq:stack_pred} 
\end{equation}

where the weights satisfy $w_k(\mathbf{x})\geq0$ and $\sum_k w_k(\mathbf{x})=1$.

The weights are inferred from the out-of-sample predictive performance of the level-0 models using Pareto-smoothed importance-sampling leave-one-out cross-validation (PSIS--LOO; \citealt{loo}). Leave-one-out cross-validation assesses how well a model predicts unseen data by repeatedly leaving out one observation and evaluating its prediction from the remaining data. PSIS--LOO provides an efficient approximation to this procedure without requiring the models to be refitted for every omitted observation. BHS uses these pointwise predictive performances to determine how much weight should be assigned to each level-0 model for every prediction. 

Unlike standard stacking, which assigns one global weight to each model, BHS allows these weights to vary smoothly with the stellar parameters. This enables the ensemble to favour different models in different regions of the parameter space while regularising the variation to reduce overfitting.

Uncertainty is propagated from the posterior predictive distributions of the three level-0 models and from the posterior distribution of the stacking weights. Predictions at a new input $\mathbf{x}_\ast$ are obtained by marginalising over this weight posterior. Further details of the stacking parametrisation, priors, and inference procedure are given in Appendix~\ref{app:bhs}. 

Bayesian hierarchical stacking has previously been shown to attain higher mean test log predictive density than fixed-weight stacking or selection of a single model in several examples presented in \citet{bhs}. We next describe each level-0 model.

\subsection{Bayesian additive regression trees}
\label{sec:bart}

BART represents the relation between the stellar observables and the target as the sum of many shallow regression trees \citep{bart}. Individual trees describe simple local behaviour, while their sum can represent non-linear relations and interactions without imposing a single global functional form. We include BART because this locally adaptive, piecewise-constant representation differs from both the smooth covariance assumption of the TP-HSGP and the global non-linear mapping of the HBNN.

BART estimates an unknown function \( f \) that maps a \(p\)-dimensional input vector \( \mathbf{x} \) to a response \( y \):
\begin{subequations}\label{eq:bart1}
\begin{align}
    y &= f(\mathbf{x}) + \epsilon,\\
    \epsilon &\sim \mathcal{N}(0, \sigma^2),
\end{align}
\end{subequations}
where $\epsilon$ represents the residual noise and $\sigma$ is its standard deviation. The same residual scale appears in the complete BART likelihood given in Appendix \ref{app:bart}.

The regression function $f(\mathbf{x})$ is approximated by an ensemble of $m$ regression trees,
\begin{equation}
    f(\mathbf{x}) \approx \sum_{j=1}^{m} g_j(\mathbf{x}),
    \label{eq:sumtree}
\end{equation}
where each tree $g_j$ recursively partitions the predictor space using binary decision rules of the form $x_d < c$. Each terminal node of a tree is associated with a constant parameter inferred from the observations assigned to that node, and the prediction of the tree for a given input corresponds to the parameter of the terminal node into which the input falls.

We use $m=250$ trees. The tree-depth prior and the remaining implementation-specific priors are described in Appendix \ref{app:bart}.

To propagate measurement uncertainties in the predictors, we represent the noise-free input vector $\mathbf{x}_i$ as a latent quantity conditioned on the measured values:

\begin{equation}
\mathbf{x}_i \mid \tilde{\mathbf{x}}_i,\mathbf{s}_i
\sim
\mathcal{N}\!\big(
\tilde{\mathbf{x}}_i,\mathrm{diag}(\mathbf{s}_i^2)
\big),
\label{eq:eiv}
\end{equation}
where $\tilde{\mathbf{x}}_i = (T_{\mathrm{eff}}, L, \log g, [\mathrm{Fe/H}])_i$ are the measured stellar parameters and $\mathbf{s}_i$ encodes their reported uncertainties. No augmented training sample is generated; the latent inputs are instead marginalised over during posterior inference.

\subsection{Two-process heteroscedastic sparse Gaussian process}

A Gaussian process describes a distribution over smooth functions, with a covariance kernel determining how strongly the function values at two input locations are related \citep{mackay1998,rasmussen2006gpml}. In the present application, stars with similar stellar parameters are therefore expected to have similar masses or radii, with the characteristic scale of this similarity inferred from the data. We include a GP in the ensemble because this explicit smoothness assumption is complementary to the local tree partitions of BART and the more flexible global mapping of the HBNN.

We model the stellar parameter of interest (mass or radius) \(y\) using a two-process heteroscedastic GP. The first process \(f(\mathbf{x})\) models the predictive mean as a GP over the observed stellar parameters \(\mathbf{x}\), while the second process \(g(\mathbf{e})\) models the log noise variance as a GP over the reported measurement errors \(\mathbf{e}\).  
The likelihood for the observed quantity is therefore
\begin{equation}
    y \sim \mathcal{N}\left(f(\mathbf{x}), \exp\left(\frac{1}{2} \times g(\mathbf{e})\right)\right).
\end{equation}

Here, $\mathbf{x}=(T_{\mathrm{eff}},\log g,[\mathrm{Fe/H}],L)$ and $\mathbf{e}=(\sigma_{T_{\mathrm{eff}}}, \sigma_{\log g},\sigma_{[\mathrm{Fe/H}]}, \sigma_L)$.

Separating the two processes allows the mean relation and its residual variance to be modelled separately. A single homoscedastic likelihood would assign the same residual variance to all stars, despite the substantial differences between their reported
measurement uncertainties shown in Fig. \ref{fig:uncert}. The process $g(\mathbf{e})$ allows the predictive variance to depend on these differences without forcing the mean process $f(\mathbf{x})$ to interpret changes in observational uncertainty as structure in the stellar relation.

For both processes, we use an automatic relevance determination (ARD) squared-exponential kernel:
\begin{equation}
k(\mathbf{x}, \mathbf{x'})
=
\prod_{d=1}^{4} \eta_d^2 \exp\!\left(-\frac{(x_d-x_d')^2}{2 \ell_d^2}\right),
\label{eq:post}
\end{equation}
where each input dimension $d$ has its own length-scale $\ell_d$, and $\eta_d$ is the signal amplitude associated with that dimension. We additionally include a small white-noise jitter term for numerical stability.

Both processes are implemented using $M$ inducing points to reduce the computational cost, and their kernel hyperparameters and inducing variables are inferred jointly rather than fixed at point estimates. The complete sparse-GP specification and hyperpriors are given in Appendix \ref{app:gp}.

\subsection{Hierarchical Bayesian neural network}

\begin{figure}
    \centering
   \includegraphics[width=\columnwidth]{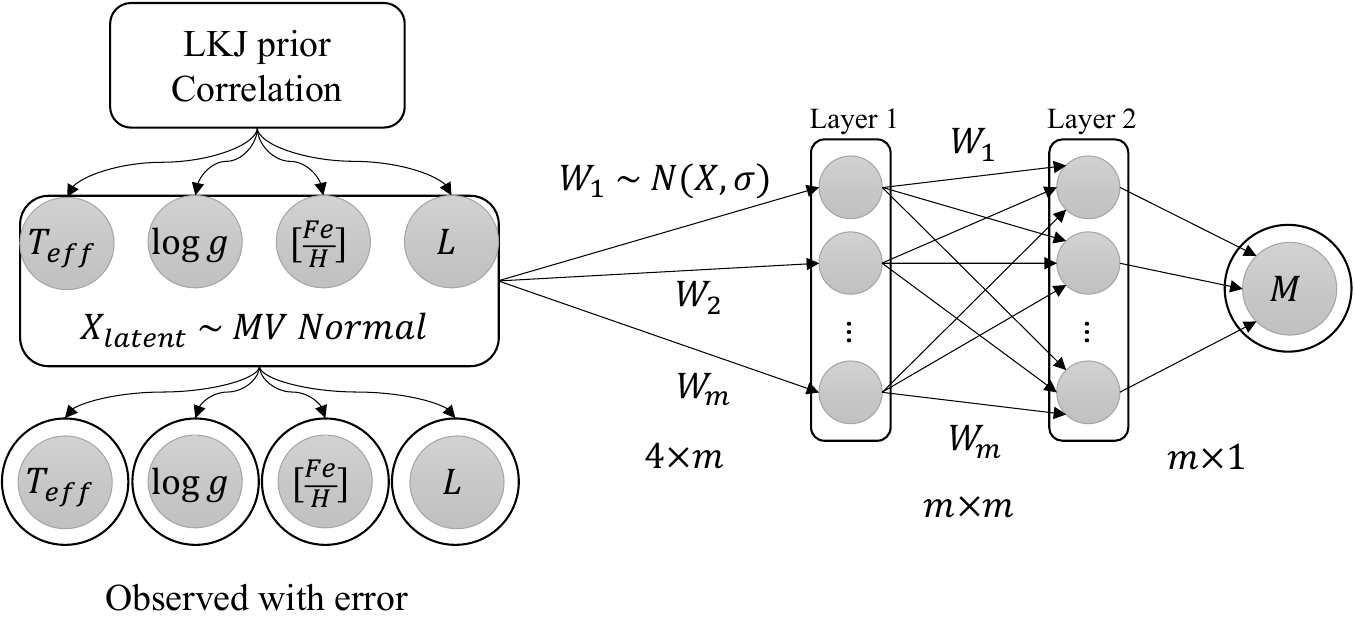} 
   \caption{Logical structure of the hierarchical Bayesian neural network model. The first level is for the model parameters: the latent stellar parameters and the neural networks for which we used 2 hidden layers each activated by leaky ReLU functions. The second level is for the hyperparameters: the covariance between parameters learnt from the data. $m$ corresponds to the number of nodes of the neural network.}
   \label{fig:hbnn}
\end{figure}

The HBNN provides a flexible global non-linear mapping from the stellar observables to mass or radius, complementing the smooth TP-HSGP and the local partitions of BART. It represents the measured inputs as noisy observations of correlated latent quantities, thereby propagating both measurement uncertainties and inter-parameter correlations through the neural network as seen in Fig. \ref{fig:hbnn}.

The covariance among $T_{\mathrm{eff}}$, luminosity, $\log g$, and metallicity is regularised using an LKJ prior \citep{LKJ_2009}. This is a statistical regularisation choice that ensures a valid correlation matrix without imposing a particular stellar-physics relation among the inputs. Its mathematical specification is given in Appendix~\ref{app:hbnn}.

The latent inputs are passed through two hidden layers with leaky-ReLU activations. This compact architecture provides a flexible non-linear mapping while limiting model complexity for the size of the training sample. Gaussian priors are assigned to the network weights and biases. 

We use a heavy-tailed Student-$t$ likelihood for stellar mass. With a Gaussian likelihood, the mass model exhibited MCMC convergence problems, shown by elevated $\hat{R}$ values, together with multiple high Pareto-$k$ values in the PSIS--LOO diagnostics, indicating sensitivity to influential observations. The Student-$t$ likelihood reduced their influence and resolved these diagnostic problems. The corresponding problems did not arise for radius, for which we therefore retain a Gaussian likelihood. The latent-input model, network priors, and target likelihoods are specified in Appendix~\ref{app:hbnn}. A similar HBNN architecture was recently applied to another astrophysical problem by \citet{tamames_2025}, where the model is described in greater detail.

\section{Results}

\subsection{Evaluation metrics}
\label{sec:eval_met}

We evaluate three complementary aspects of the predictions. The mean absolute error (MAE), mean absolute relative difference (MARD), and mean relative difference (MRD) quantify the agreement of point estimates with the adopted reference values. The log predictive density (LPD) and continuous ranked probability score (CRPS) evaluate the complete posterior predictive distributions, accounting for both their location and concentration. Finally, empirical coverage and normalised residuals assess whether the reported predictive uncertainties are calibrated with respect to the reference values.

\subsubsection{Point-prediction metrics}
\label{sec:pp_met}

Let $t_i$ denote the adopted literature reference value for star $i$, $\hat{t}_i$ the corresponding point prediction, and $N$ the number of stars in the evaluation sample. In this work, $\hat{t}_i$ is the posterior predictive mean.

The mean absolute error is 
\begin{equation} 
\mathrm{MAE} = \frac{1}{N} \sum_{i=1}^{N} \left| \hat{t}_i-t_i \right|.
\label{eq:MAE} 
\end{equation}

The MAE expresses the typical absolute difference in the physical units of the target, $M_\odot$ or $R_\odot$. It is therefore useful for the direct comparison with the BASTA mass estimates in Sect.~\ref{sec:basta}, where the physical scale of the differences is of interest.

To compare fractional differences across targets with different masses or radii, we use the mean absolute relative difference,
\begin{equation}
    \text{MARD} = \frac{\sum_{i=1}^N \frac{|t_i- \hat{t_i}|}{t_i}}{N} \times 100 \;,
    \label{eq:MARD}
\end{equation}
and the mean relative difference,
\begin{equation}
    \text{MRD} = \frac{\sum_{i=1}^N \frac{t_i- \hat{t_i}}{t_i}}{N} \times 100 \;.
    \label{eq:MRD}
\end{equation}

The MARD measures the mean absolute fractional difference from the reference values, irrespective of direction, whereas the MRD measures the mean signed fractional offset. With the convention in Eq.~\ref{eq:MRD}, a positive MRD indicates that the predictions are, on average, smaller than the reference values. These quantities measure agreement with the adopted literature values and not absolute astrophysical accuracy.

\subsubsection{Probabilistic metrics}
\label{sec:probmet}

Point-prediction metrics do not assess whether the width and shape of a predictive distribution appropriately represent its uncertainty. We therefore also evaluate the posterior predictive distributions using proper scoring rules \citep{gneiting2007strictly, gelman2013bayesian}. 

For each star, we combine the model predictive distribution with the reported uncertainty of its literature reference value. Let $p_{\mathrm{pred},i}(y\mid\mathbf{x}_i,\mathcal{D})$ be the model posterior predictive density, where $\mathcal{D}$ is the training dataset, and let $s_i$ be the reported $1\sigma$ uncertainty of the reference value $t_i$. Assuming that the predictive and reference uncertainties are independent and approximating the latter as Gaussian, we define the combined evaluation density as \begin{equation} 
p_{\mathrm{comb},i}(u) = \int_{-\infty}^{\infty} p_{\mathrm{pred},i} (y\mid\mathbf{x}_i,\mathcal{D}) \, \mathcal{N}(u\mid y,s_i^2) \,\mathrm{d}y. 
\label{eq:combined_predictive} 
\end{equation}
Here, $y$ is a possible model-predicted value and $u$ is a possible value after including the reference uncertainty.
Thus, the probabilistic metrics account for both uncertainty in the model prediction and uncertainty in the adopted reference value.

The mean log predictive density is
\begin{equation}
\mathrm{LPD} = \frac{1}{N} \sum_{i=1}^{N} \log p_{\mathrm{comb},i}(t_i).
\label{eq:lpd} 
\end{equation}
The LPD rewards predictive distributions that assign high probability density to the reference values; higher values indicate better probabilistic predictive performance.

The continuous ranked probability score is defined as 
\begin{equation} 
\mathrm{CRPS} = \frac{1}{N} \sum_{i=1}^{N} \int_{-\infty}^{\infty} \left[ F_{\mathrm{comb},i}(u) - \mathbb{I}(u\geq t_i) \right]^2 \,\mathrm{d}u, 
\label{eq:crps} 
\end{equation}
where $F_{\mathrm{comb},i}$ is the cumulative distribution function corresponding to Eq.~\ref{eq:combined_predictive} and $\mathbb{I}$ is the indicator function. Unlike the LPD, which evaluates the density at the reference value, the CRPS compares the complete cumulative predictive distribution with that value. It therefore evaluates both the location and concentration of the distribution. Lower values indicate better probabilistic predictive performance and the units are the same as the predicted quantity.

We further assess calibration using the empirical coverage of the combined predictive intervals and the normalised residuals. Let $q_{0.16,i}$ and $q_{0.84,i}$ denote the 16th and 84th percentiles of $F_{\mathrm{comb},i}$. The empirical 68 per cent coverage is
\begin{equation} 
C_{68} = \frac{100}{N} \sum_{i=1}^{N} \mathbb{I} \left( q_{0.16,i} \leq t_i \leq q_{0.84,i} \right). 
\label{eq:coverage} 
\end{equation}
For calibrated predictive distributions, this quantity should be close to 68 per cent.

For the normalised residuals, we define 
\begin{equation} 
\sigma_{\mathrm{tot},i} = \sqrt{ \sigma_{\mathrm{pred},i}^{2} + s_i^{2} },
\label{eq:sigma_total} 
\end{equation} 
where $\sigma_{\mathrm{pred},i}$ is the standard deviation of the model posterior predictive distribution. 
The normalised residual is then 
\begin{equation} 
z_i = \frac{t_i-\hat{t}_i} {\sigma_{\mathrm{tot},i}},
\label{eq:normalised_residual} 
\end{equation} 
where $\hat{t}_i$ is the corresponding posterior predictive point estimate. For calibrated, approximately Gaussian uncertainties, the distribution of $z_i$ should have a mean close to zero and a standard deviation $\sigma_z$ close to unity. Values of $\sigma_z<1$ indicate that the combined uncertainties are wider than required by the observed residuals, whereas $\sigma_z>1$ indicates that they are too narrow.

\subsection{Test set analysis}

We start by reviewing the performance of our models on the test set which represents 20 per cent of our data described in Sect. \ref{sec:data} and shown in Fig. \ref{fig:traintest}.

\subsubsection{Model comparison and evaluation}

\begin{figure*}
    \centering
    \includegraphics[width=\textwidth]{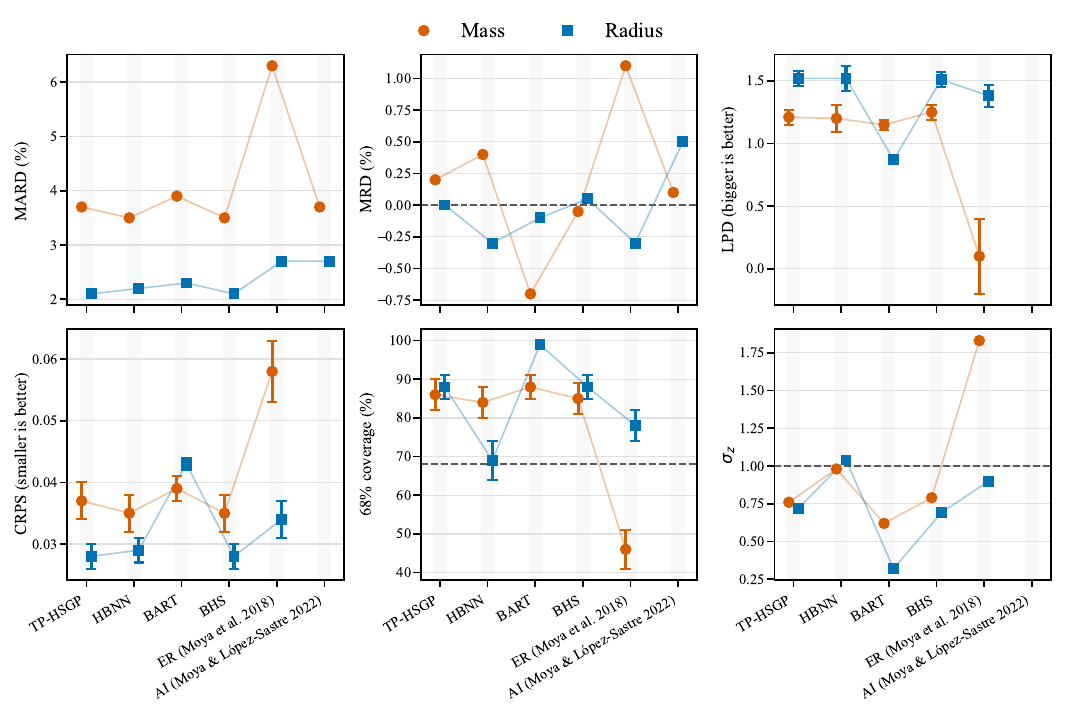}
    \caption{
    Performance on the independent test set for stellar mass (orange circles) and radius (blue squares). Arrows indicate the preferred metric direction. Error bars denote the standard error of the mean for LPD and CRPS and the binomial standard error for coverage. Dashed lines mark MRD$=0$, coverage$=68$ per cent, and $\sigma_z=1$. ER denotes the empirical relations and AI the frequentist stacking model; AI is absent from the probabilistic panels because it does not provide predictive distributions.
    }
    \label{fig:model_metrics}
\end{figure*}

Fig. \ref{fig:model_metrics} reports the metrics of all our models (with and without stacking), as well as a comparison with the state-of-the-art models in \citet{moya1} and  \citet{moya2} for results on the same test set. Note that while \citet{moya1}'s approach is based on the use of empirical relations, in \citet{moya2} a stacking model was also used. Numerical values for all these metrics can be found in Appendix \ref{app:metrics}.

In this section we focus on the point-prediction metrics, MARD and MRD, presented in Fig. \ref{fig:model_metrics} and defined in Sect. \ref{sec:pp_met}. In the case of stellar mass, the HBNN has the lowest MARD while the TP-HSGP has the smallest MRD among our individual models. The BHS matches the HBNN for MARD and outperforms the TP-HSGP for MRD highlighting it as having the best agreement with the test set.

\begin{figure*}
   \centering
     \begin{subfigure}[b]{1.0\textwidth}
         \centering
         \includegraphics[width=\textwidth]{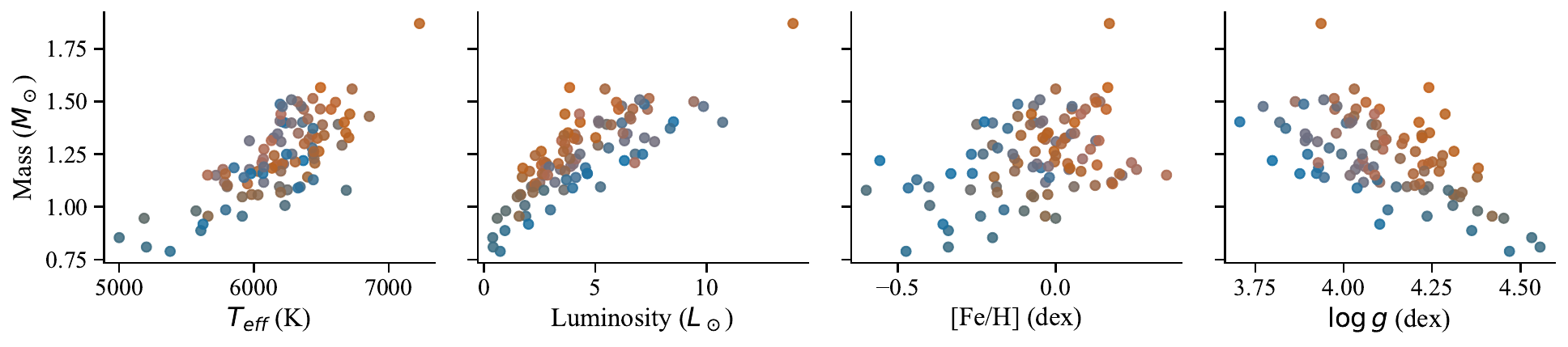}
         \caption{Mass}
         \label{fig:weight_mass}
     \end{subfigure}
     \hfill
     \begin{subfigure}[b]{1.0\textwidth}
         \centering
         \includegraphics[width=\textwidth]{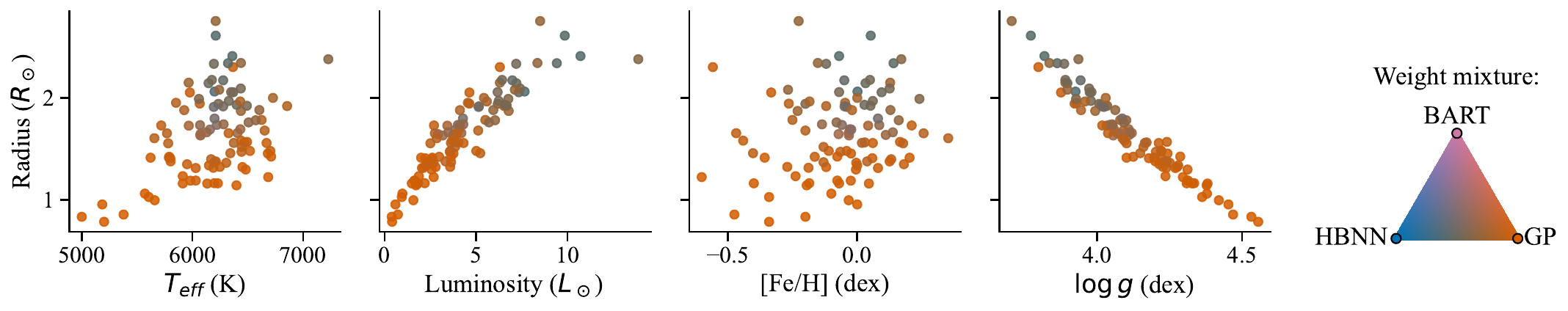}
         \caption{Radius}
         \label{fig:weight_radius}
     \end{subfigure}
     \caption{Weights (hue) assigned to each model by BHS across parameter space for predictions on the test set.}
     \label{fig:weights}
\end{figure*}

Fig. \ref{fig:weight_mass} gives us insight into the weight attributed by BHS to each model across the parameter space and therefore their strengths and weaknesses. The first point we notice is that BART rarely dominates. We can attribute this to its simplistic assumption of independent input parameters. 

We also notice clear regions of TP-HSGP or HBNN dominance. The HBNN performs best in more extreme regions such as low $T_\text{eff}$ or high gravity where less training data are available, whereas the TP-HSGP dominates in the more monotonic, well-behaved areas of parameter space. This can be explained by the more complex mapping offered by a neural network. This complexity is not captured as well by the smooth kernels (covariance functions) of the TP-HSGP.

Now coming back to Fig. \ref{fig:model_metrics}, the TP-HSGP has the lowest MARD and MRD for radius estimates. Again, we can use a similar argument as in the previous paragraph to explain the preference for mass and radius by the HBNN and TP-HSGP respectively. Radius has a more linear relationship with our input parameters than mass; it is more evenly spread out over this space as seen in Fig. \ref{fig:weights}. We also notice a stronger TP-HSGP dominance in the weights for radius. 

In Fig. \ref{fig:model_metrics}, we also compare our results to the best empirical relation (ER) found in \cite{moya1} and a stacking algorithm combining six frequentist machine-learning techniques \citep[AI,][]{moya2} on the same test set. We can conclude that by propagating uncertainties and including input parameter correlations we achieve predictions that show closer agreement with the adopted reference values and better-calibrated uncertainties.

The BHS test-set results are shown in Appendix Fig.~\ref{fig:test}. They show no pronounced residual structure for radius, while the mass predictions suggest a mild underestimation between approximately $1.4$ and $1.6\,M_\odot$.

\subsubsection{Probabilistic evaluation}

While the MARD and MRD compare the point-prediction performance of the different methods, they do not assess the quality of the associated predictive uncertainties. To evaluate the full posterior predictive distributions, we employ the mean log predictive density (LPD) and the continuous ranked probability score (CRPS) as described in Sect. \ref{sec:eval_met}.

We now focus on the probabilistic metrics presented in Fig. \ref{fig:model_metrics}.
The uncertainties shown for the mean LPD and CRPS are standard errors of the mean over the test set, while the coverage uncertainty is the corresponding binomial standard error.
For stellar mass, the Bayesian hierarchical stacking (BHS) model achieves the highest mean LPD, indicating the best overall probabilistic performance on the test set. However, to compare two models, one takes the mean difference in LPD between the two. Relative to the best individual model, the TP-HSGP, we obtain
\[
\Delta \mathrm{LPD}_{\mathrm{BHS-\text{TP-HSGP}}} = 0.034 \pm 0.024,
\]
which is larger than its standard error, although the improvement remains modest. The BHS also attains the lowest CRPS, tied with the HBNN, while substantially outperforming the empirical relations in all probabilistic metrics. The empirical relations exhibit both a much lower LPD and severe undercoverage, with only 46 per cent of targets falling within the nominal 68 per cent interval and $\sigma_z = 1.83$, indicating strongly underestimated uncertainties.

For stellar radius, all three Bayesian models achieve nearly identical LPD and CRPS values. The mean differences in LPD between the BHS and the best individual models are smaller than their corresponding standard errors, indicating that the models are statistically indistinguishable in terms of probabilistic performance. This behaviour is consistent with the simpler and more monotonic dependence of stellar radius on the input parameters. As a result, there is less model diversity for the stacking procedure to exploit, and combining models yields little additional benefit.

In contrast, stellar mass presents a more complex mapping from the atmospheric parameters to the target quantity, particularly in sparsely sampled regions of parameter space where the HBNN and TP-HSGP exhibit complementary strengths. In this case, the BHS is able to combine the advantages of both models and achieve a modest but consistent improvement in predictive performance.

The calibration diagnostics show that most of our Bayesian models are conservative, with empirical 68 per cent coverages significantly above the nominal value and $\sigma_z < 1$. This indicates that the predictive intervals are somewhat wider than strictly necessary. In astrophysical applications, however, conservative uncertainties are preferable to overconfident ones, particularly when inferred stellar parameters are propagated into subsequent analyses such as exoplanet mass and radius determinations. Moreover, because the observational uncertainties of the test set are included in the evaluation, some degree of overcoverage is expected. The radius HBNN is the best-calibrated model overall, with $\sigma_z$ values closest to unity.

Overall, these results demonstrate that the BHS provides the most reliable probabilistic predictions for stellar mass and performs comparably to the best individual models for stellar radius on the held-out set. More broadly, they show that Bayesian machine-learning models not only improve point prediction performance relative to empirical relations, but also deliver safer uncertainty estimates. Having established the performance of the full four-parameter models, we now investigate how these results change when not all input parameters are available.

\subsubsection{Performance with reduced input parameters}

It is important to note that the individual Bayesian models can accommodate missing input parameters at prediction time by marginalising over the corresponding latent variables. In contrast, the Bayesian hierarchical stacking (BHS) model combines predictions from models trained on a fixed set of inputs and therefore requires all parameters used during training to be available. Consequently, when one or more input parameters are systematically absent, the stacking model must be retrained using only the subset of parameters that is consistently available.

In the context of this work, and in many stellar catalogues, luminosity is frequently unavailable. We therefore retrain all models using only $T_{\mathrm{eff}}$, [Fe/H], and $\log g$, and evaluate their performance in this reduced-input setting.

We find that retraining the models with the available parameters yields more accurate predictions with regard to the test set than applying four-parameter models with luminosity treated as missing. For example, the HBNN mass predictions improve from a MARD of 4.4 per cent when using the four-parameter model without luminosity to 3.7 per cent when the model is retrained directly on the three available inputs. Similar improvements are observed for radius. These results indicate that, when luminosity is the only missing parameter, retraining on the reduced set of inputs can provide a measurable improvement in predictive performance rather than treating this parameter as missing in a 4-input model. 

\begin{table}
    \centering
    \caption{
    Performance of models trained on three input parameters ($T_{\mathrm{eff}}$, $[\mathrm{Fe/H}]$, and $\log g$) on the held-out test set.
    }
    \footnotesize
    \setlength{\tabcolsep}{3pt}
    \resizebox{\columnwidth}{!}{
    \begin{tabular}{l|cc|cc}
        \toprule
        &
        \multicolumn{2}{c|}{Mass} &
        \multicolumn{2}{c}{Radius} \\
        \cmidrule(r){2-3}
        \cmidrule(l){4-5}

        Model &
        MARD (\%) &
        LPD $\uparrow$ &
        MARD (\%) &
        LPD $\uparrow$ \\
        \midrule

        TP-HSGP &
        3.6 &
        $1.19 \pm 0.06$ &
        1.8 &
        $1.50 \pm 0.07$ \\

        HBNN &
        3.7 &
        $1.15 \pm 0.1$ &
        2.1 &
        $1.51 \pm 0.09$ \\

        BART &
        3.9 &
        $1.05 \pm 0.03$ &
        2.2 &
        $0.86 \pm 0.01$ \\

        BHS &
        3.5 &
        $1.22 \pm 0.06$ &
        1.8 &
        $1.51 \pm 0.05$ \\

        ER {\tiny \citep{moya1}} &
        6.6 &
        $-0.3 \pm 0.5$ &
        2.4 &
        $1.43 \pm 0.05$ \\

        \bottomrule
    \end{tabular}
    }
    \tablefoot{Representative of the common case where luminosity is unavailable. The MARD and mean LPD are used as described in Sect. \ref{sec:eval_met} and Fig. \ref{fig:model_metrics}.}
    \label{tab:perf_3param}
\end{table}

Table \ref{tab:perf_3param} summarises the performance of models retrained using only $T_{\mathrm{eff}}$, [Fe/H], and $\log g$. Overall, the predictive performance on the test set remains very similar to that of the full four-parameter models, showing that reliable stellar mass and radius estimates can still be obtained when luminosity is unavailable.

For stellar mass, the BHS retains the best overall probabilistic performance on the test set, achieving the highest mean LPD ($1.22 \pm 0.06$), while maintaining the same MARD (3.5 per cent) as in the four-parameter case. This indicates that the advantages of combining models persist even when luminosity is omitted from the input set.

For stellar radius, the results are similarly strong. The TP-HSGP and BHS achieve the lowest MARD (1.8 per cent), which is slightly lower than in the four-parameter case, while the TP-HSGP, HBNN, and BHS all have statistically indistinguishable LPD values. This suggests that removing luminosity does not degrade the predictive performance on the test set for radius and may even lead to a small improvement in point agreement, likely because the reduced model avoids incorporating the partially redundant information coming from $\log g$ and luminosity.

These results show that removing luminosity has little impact on predictive performance. Reliable stellar mass and radius estimates can therefore be obtained using only $T_{\mathrm{eff}}$, [Fe/H], and $\log g$, which is particularly useful for exoplanet host stars where luminosity is often unavailable.

\subsection{Comparison with BASTA}
\label{sec:basta}

To compare our data-driven approach with conventional grid-based stellar modelling, we use the Kepler LEGACY sample analysed by \citet{silva_2017}. This study reports stellar properties obtained with several asteroseismic modelling pipelines, including BASTA \citep{basta}. Of the 66 stars in the original sample, 22 are present in the sample used in this work. We exclude these objects, leaving 44 stars.

For these 44 stars, we estimate the stellar masses using BHS and the empirical relations (ER) of \citet{moya1}. Both methods use the same three input parameters: $T_{\rm eff}$, $[\mathrm{Fe/H}]$, and $\log g$, taken from \citet{silva_2017}. We use the three-parameter version of BHS and omit luminosity because both luminosity and $\log g$ are derived from the BASTA analysis. This choice limits the dependence of our predictions on the BASTA results used for comparison, although the dependence is not removed entirely because the adopted $\log g$ values are retained. We then compare the resulting mass estimates with the BASTA masses reported by \citet{silva_2017}. BASTA combines stellar-evolution grids with spectroscopic and asteroseismic constraints, whereas BHS and the empirical relations use only the three parameters listed above. Therefore, this comparison evaluates the agreement between methods with different modelling assumptions and input requirements.

\begin{figure}
    \centering
    \includegraphics[width=\columnwidth]{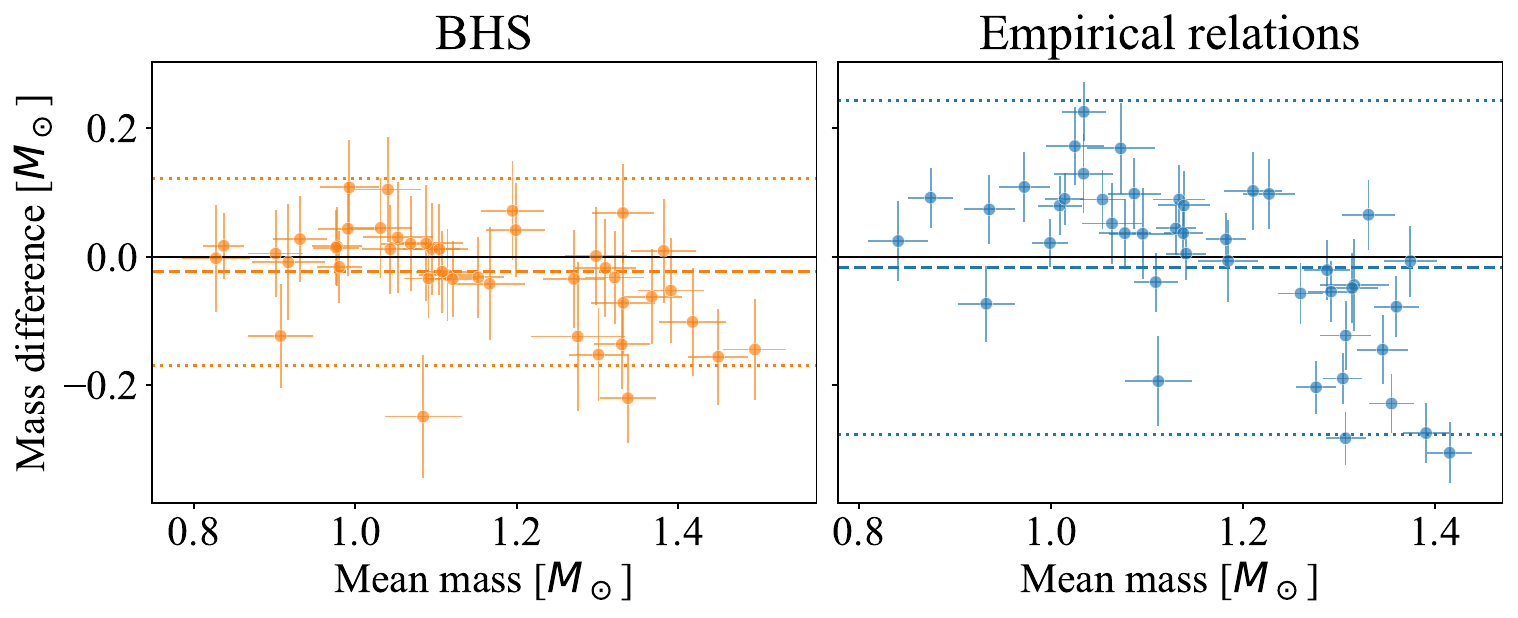} 
    \caption{Masses obtained with BHS and empirical relations \citep{moya1} compared to those estimated with BASTA \citep{basta}. We present the $\text{method}-\text{BASTA}$ difference against the mean of the two estimates. Error bars show the uncertainties in the differences, obtained by adding the quoted uncertainties in quadrature. The dashed lines correspond to the mean of the differences weighted by their uncertainties while the dotted lines correspond to twice the weighted standard deviation around the mean.  Both BHS and the empirical relations use $T_{\rm eff}$, $[\mathrm{Fe/H}]$, and $\log g$ from \citet{silva_2017}.}
    \label{fig:basta}
\end{figure}

\begin{table}
    \caption{Pairwise comparison of BHS, empirical-relation (ER), and BASTA mass estimates.}
    \label{tab:basta}
    \centering
    \begin{tabular}{lccc}
        \toprule
        Metric & BHS--BASTA & BHS--ER & ER--BASTA \\
        \midrule
        MAE ($M_\odot$) & 0.058 & 0.059 & 0.100 \\
        MARD (\%)       & 4.8   & 5.2   & 8.5   \\
        MRD (\%)        & 1.8  & 2.0  & -0.8   \\
        \bottomrule
    \end{tabular}
    \tablefoot{The sample comprises 44 Kepler LEGACY stars \citep{silva_2017}.}
\end{table}

In Fig. \ref{fig:basta}, for each method, we present the difference $M_{\rm method}-M_{\rm BASTA}$ against the mean of the two mass estimates. The BHS estimates show closer agreement with BASTA than the empirical relations do, as indicated by the MAE and MARD values in Table~\ref{tab:basta}. 
The BASTA--BHS comparison yields a MAE of $0.058\,M_\odot$ and a MARD of 4.8 per cent, compared with $0.100\,M_\odot$ and 8.5 per cent, respectively, for BASTA--ER. 
The dispersion of the BHS--BASTA differences in Fig.~\ref{fig:basta} is also smaller than that of the ER--BASTA differences.

The BASTA--BHS MRD of $1.8$ per cent indicates a small mean offset towards lower BHS masses relative to BASTA. The empirical relations have a smaller signed offset relative to BASTA, but substantially larger absolute differences and dispersion. Figure~\ref{fig:basta} also suggests that the differences become more negative towards higher masses, particularly for the empirical relations. 

These comparisons do not establish which method is closer to the true stellar masses. BASTA is itself a model-dependent inference, and the BHS and empirical-relation estimates retain some dependence on BASTA through the adopted $\log g$ values. Instead, the comparison shows that the BHS mass estimates are broadly consistent with those from an asteroseismic grid-based analysis and show closer agreement with BASTA than the empirical relations do.

\subsection{Host-star predictions}
\label{sec:host_pred}
The models presented in the previous sections were trained exclusively on the benchmark sample described in Sect. \ref{sec:data}. In this section, we apply the resulting models to a completely independent sample of host stars from the NASA Exoplanet Archive \citep{christiansen2025nasaexoplanetarchiveexoplanet}. This independent sample serves as a test of the ability of the models to generalise beyond the training and test sets used during model development.

\subsubsection{Global comparison with the literature}

We use the default parameter set recommended by the NASA archive for each star. We only include studies with at least three available parameters ($T_{\text{eff}}, [\text{Fe/H}], \log g$). Note that we made predictions with fewer parameters, but their lower precision makes these estimations less useful, and we recommend retraining models when predicting with fewer inputs.
We restrict input values to 5 per cent above the upper and 5 per cent below the lower boundary of our training set. Predictions outside these boundaries are clearly biased, with a sharp cut-off in mass and radius showing our limited extrapolation capability. More detail about the data is given in Sect. \ref{sec:host_data}.

\begin{figure}
    \centering
   \includegraphics[width=\columnwidth]{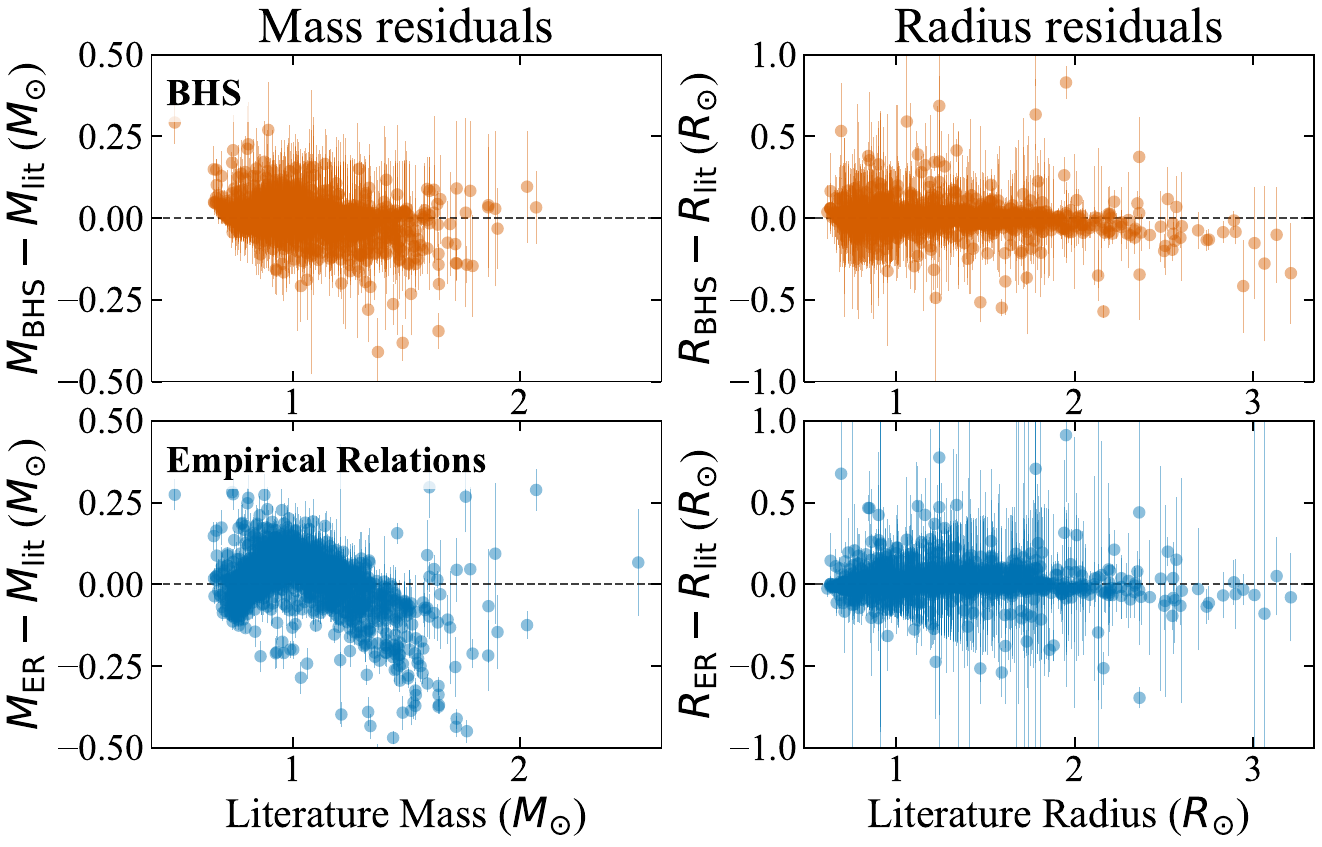} 
   \caption{Residuals between predicted and literature masses (left) and radii (right), plotted against the literature mass and radius. We show results for BHS (orange) and the empirical relations (blue) of \citet{moya1}.}
   \label{fig:full_res}
\end{figure}

\begin{table}
    \centering
    \caption{Comparison of methods using MARD and MRD.}
    \resizebox{\columnwidth}{!}{
    \begin{tabular}{lcc}
        \toprule
        Method & BHS & ER \citep{moya1}  \\
        \midrule
        Mass MARD (\%) & 2.7 & 6.2   \\
        Radius MARD (\%) & 2.7 & 3.0  \\
        Mass MRD (\%) & 0.03 & 2.7   \\
        Radius MRD (\%) & -0.001 & -0.5  \\
        \bottomrule
    \end{tabular}
    }
    \tablefoot{MARD measures agreement and MRD measures bias relative to the default literature values from the NASA Exoplanet Archive \citep{christiansen2025nasaexoplanetarchiveexoplanet}.}
    \label{tab:result_metrics}
\end{table}

Fig. \ref{fig:full_res} and Table \ref{tab:result_metrics} compare our BHS predictions and estimates from empirical relations with the literature values for 2400 host stars within the valid input range.
The BHS achieves substantially lower mean absolute relative differences (MARDs) and smaller biases (MRDs), particularly for stellar mass.
The empirical relations show a clear calibration bias: they overestimate around solar-type stars and underestimate towards cooler and hotter regimes, while the BHS predictions remain more consistent across the full range. It is important to note, however, that the literature values are themselves model-dependent estimates rather than exact ground truth. Consequently, the results presented here should be interpreted as a comparison with current catalogue values rather than as a direct measure of absolute accuracy.

\subsubsection{Star-by-star statistical consistency and outlier analysis}

To complement the distribution-level metrics in Table \ref{tab:result_metrics}, we also quantify star-by-star agreement. We classify predictions as consistent within uncertainties when $|y_{\text{BHS}}-y_{\text{lit}}| < \sqrt{\sigma_{\text{BHS}}^2 + \sigma_{\text{lit}}^2}$, and we additionally flag large discrepancies using a fractional threshold $\frac{|y_{\text{BHS}}-y_{\text{lit}}|}{y_{\text{lit}}}>10\%$. These criteria probe statistical consistency and practical significance, respectively. We find that for both mass and radius, 4.7 per cent of predictions fall outside the $1\sigma$ consistency criterion. These outliers do not show strong clustering in parameter space, nor do they preferentially originate from a single catalogue. 

Using the 10 per cent fractional threshold, 3.4 per cent of our mass predictions and 4.9 per cent of our radius predictions exceed this level. In the case of stellar mass, a single source is over-represented: approximately one quarter of the mass outliers are drawn from the sample of \citet{Christiansen_2022}, with stellar parameters derived from \citet{Ullman_2020}, which constitutes 58 per cent of the stars from this source in our sample. This behaviour can be explained by the fact that stellar masses in this catalogue are inferred from spectroscopic $\log g$, which carries large uncertainties and is strongly coupled to the adopted stellar radius. Consequently, although the fractional differences exceed 10 per cent, 85 per cent of these outliers remain consistent within the combined uncertainties. This suggests that, for these systems, our predictions provide improved agreement while remaining statistically consistent with the reported measurements.

\subsubsection{Dependence on training set coverage}

\begin{figure}
    \centering
   \includegraphics[width=\columnwidth]{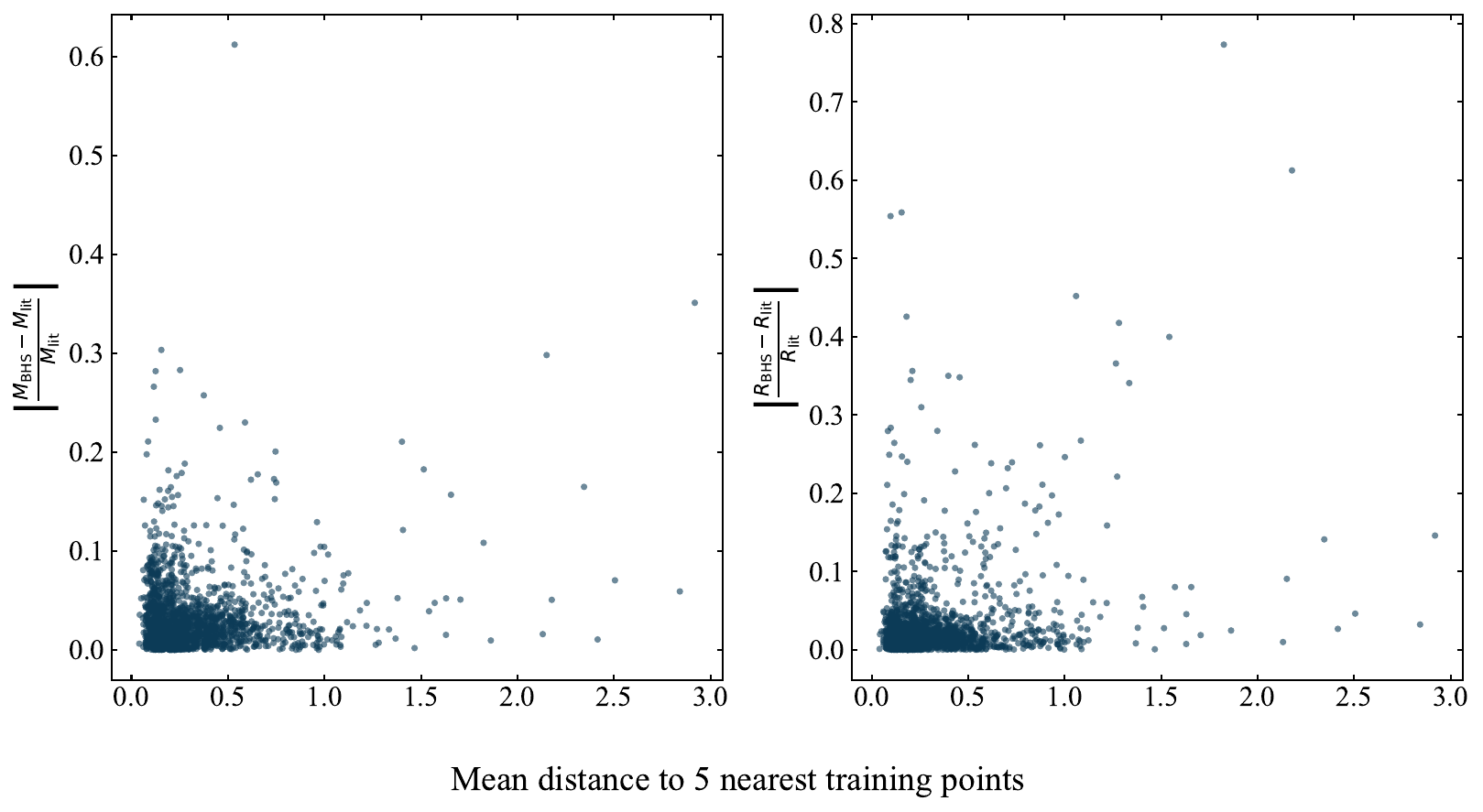} 
   \caption{Fractional residual between BHS-predicted and literature masses (left) and radii (right) against normalised average distance to the 5 closest training points in $\log g$-$T_{\text{eff}}$ space.}
   \label{fig:dist_train}
\end{figure}

To assess whether large residuals are primarily due to poor coverage in the training data, we examined the average distance to the five nearest training points in the $\log g$–$T_{\text{eff}}$ plane. Fig. \ref{fig:dist_train} shows that, although residuals increase slightly with distance, the trend is weak and highly scattered. This suggests that a significant fraction of discrepancies occur even in well-sampled regions, implying that our models may in some cases correct biases present in the literature as seen above.

Having established that residuals are not solely driven by data sparsity, we now investigate their systematic behaviour across parameter space.

\subsubsection{Residual structure across parameter space}

\begin{figure}
    \centering
   \includegraphics[width=\columnwidth]{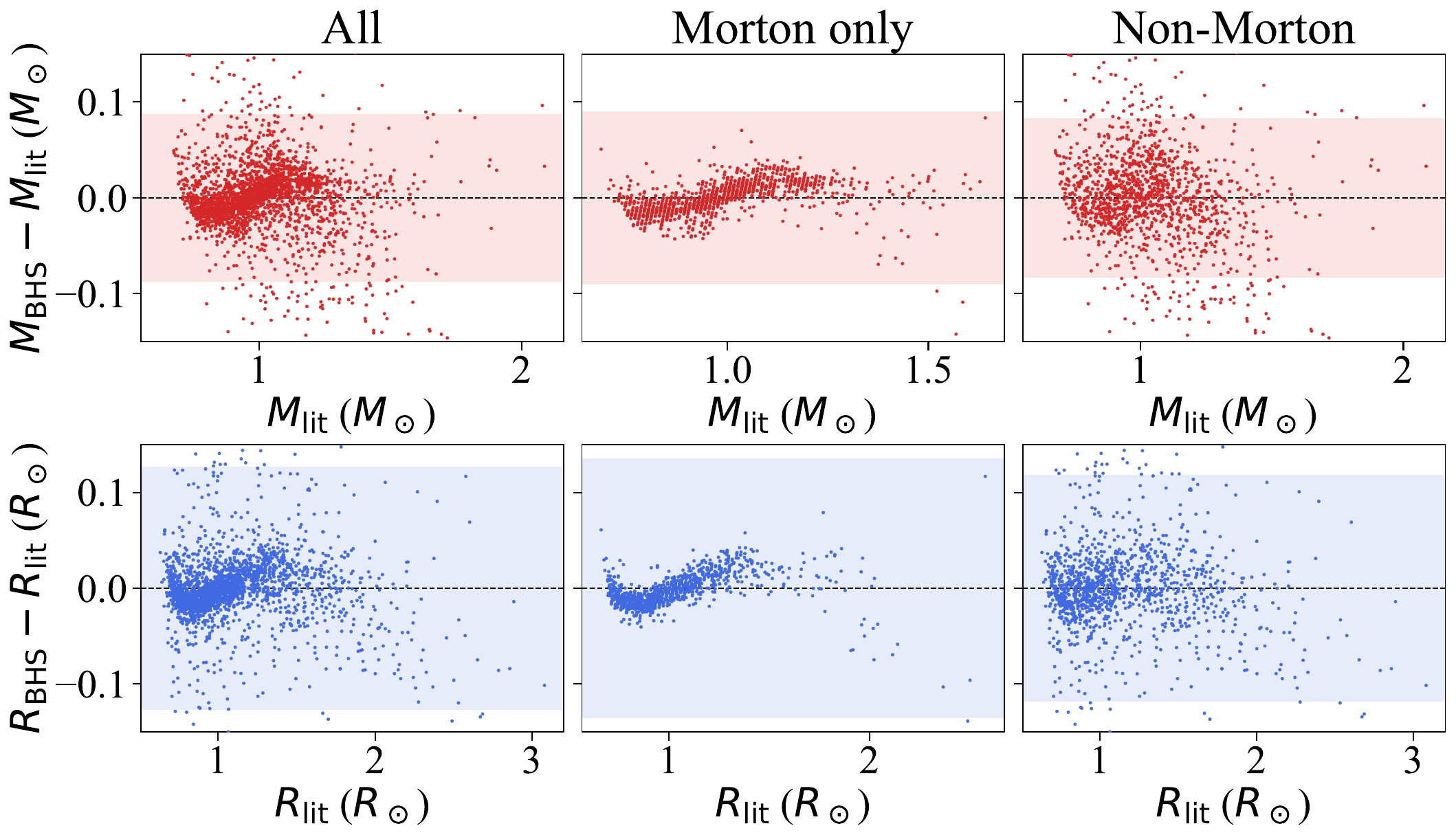} 
   \caption{Residual between BHS-predicted and literature masses (top) and radii (bottom) against mean mass or radius, respectively. We focus on smaller residuals for all values (left) and Morton et al. (2016) (centre) with the remaining stars on the right. The shaded regions indicate the median combined uncertainty of the residuals in each subset; individual uncertainties are omitted for clarity.}
   \label{fig:zoomed_res_ML}
\end{figure}
Now, we focus on smaller residual values that concern most of our predictions.
We concentrate on \cite{morton_2016} and compare our estimates of stellar mass and radius in Fig. \ref{fig:zoomed_res_ML} because it is the most represented catalogue with 1078 stars. In this figure we notice a clear bias in the residual of both mass and radius. It is important to note that we find overall good agreement with this catalogue, and the bias remains within the combined uncertainties. However, it is still an interesting artefact that we will try to unravel. 

Below $\approx$ 0.8 $M_{\odot} / R_{\odot}$, our models approach the lower boundary of the training domain.
The predictive mean saturates because the model’s learnt function flattens outside the data-supported region, effectively imposing a lower limit on predicted mass and radius. This explains why the residual increases as we go down the x-axis for both mass and radius. 

The second bias we notice is a linearly increasing one around solar masses and radii. We also find a similar bias when comparing literature with empirical relations. This indicates that the bias originates, at least in part, from correlations embedded in the training dataset, which overlaps with empirical calibrations.

\begin{figure}
    \centering
   \includegraphics[width=\columnwidth]{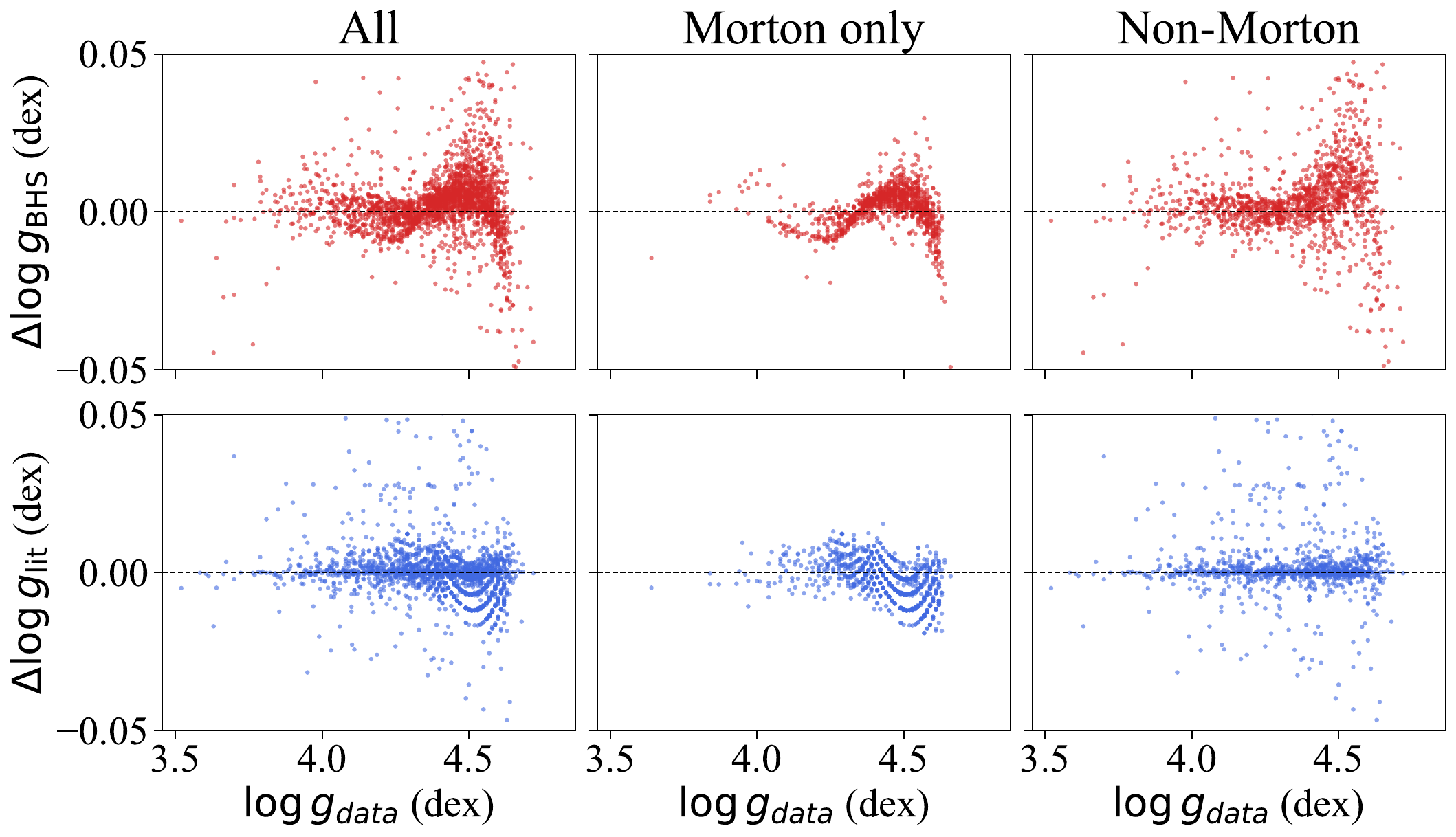} 
   \caption{Difference between $\log g$ computed using mass and radius and $\log g$ from the data as a function of the latter. At the top we compute with BHS mass and radius ($\Delta \log g_{\text{BHS}}=\log g(M_{\text{BHS}}, R_{\text{BHS}}) - \log g_{\text{data}}$) and at the bottom we use literature values ($\Delta \log g_{\text{lit}}=\log g(M_{\text{lit}}, R_{\text{lit}}) - \log g_{\text{data}}$). Note that we focus on the $[-0.05,0.05]$ window in $\Delta \log g$.}
   \label{fig:logg_comp}
\end{figure}

To get a clearer idea of where this bias originates from, we can test the theoretical coherence or internal consistency of our predictions using $\log g$ as shown in Fig. \ref{fig:logg_comp}.

\begin{figure}
   \centering
     \begin{subfigure}[b]{\columnwidth}
         \centering
         \includegraphics[width=\columnwidth]{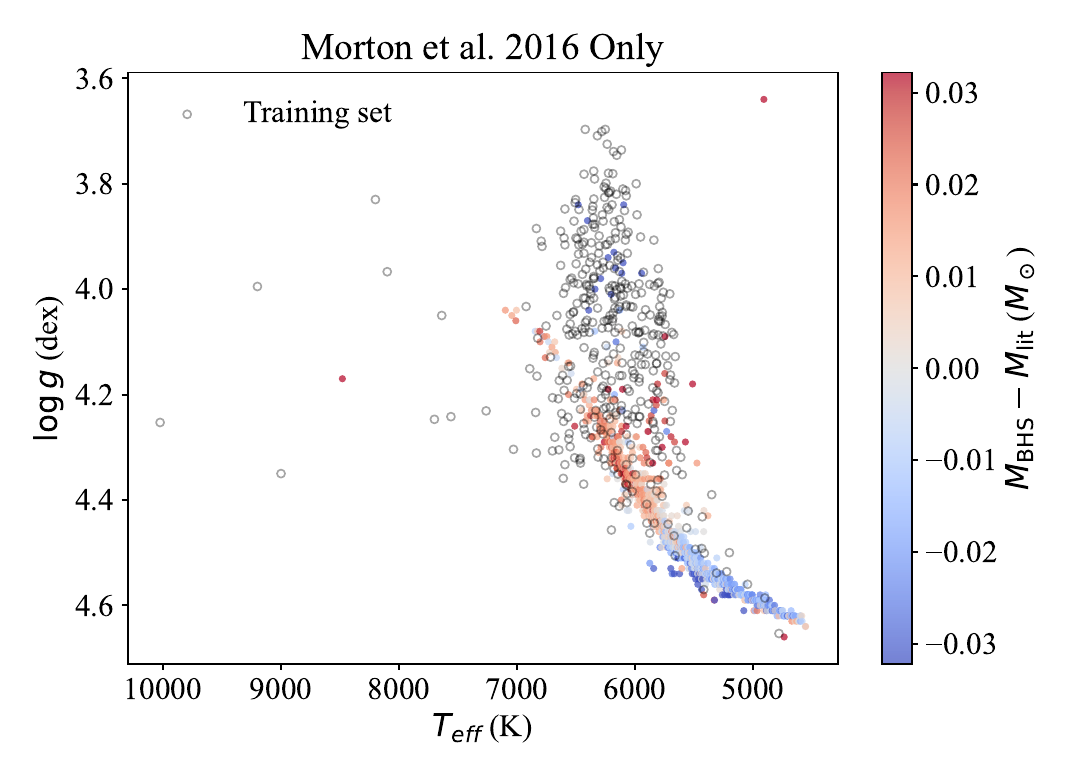}
         \caption{Mass}
         \label{fig:HR_mass}
     \end{subfigure}
     \hfill
     \begin{subfigure}[b]{\columnwidth}
         \centering
         \includegraphics[width=\columnwidth]{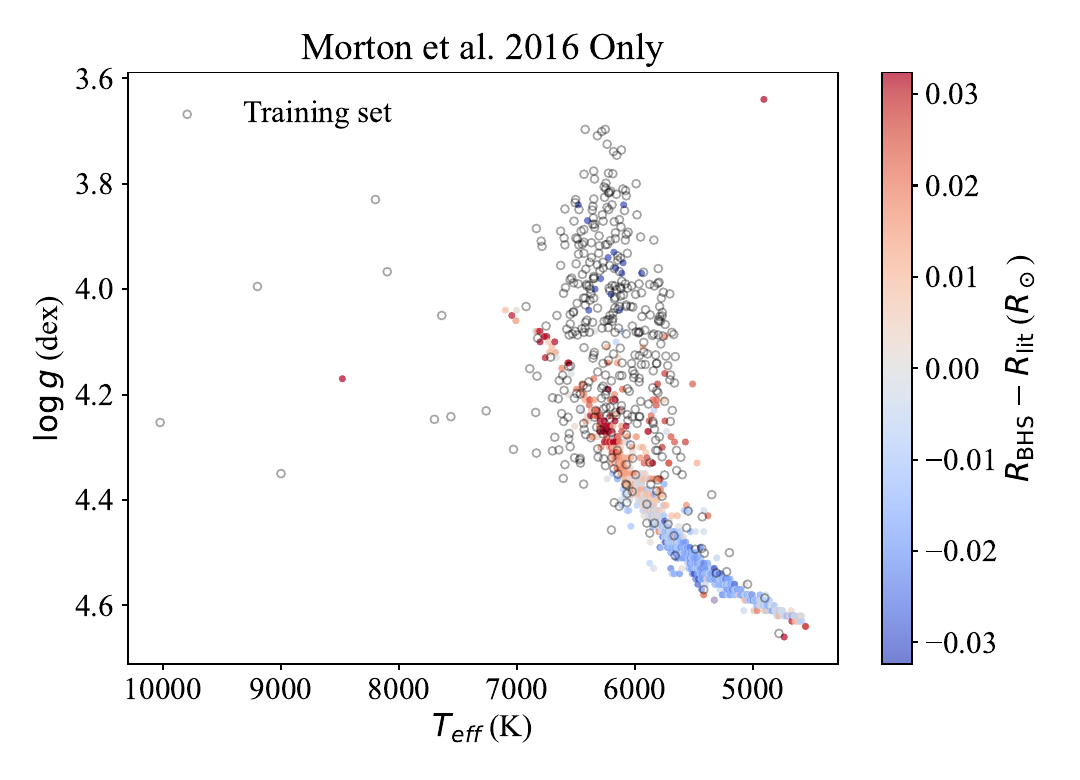}
         \caption{Radius}
         \label{fig:HR_radius}
     \end{subfigure}
     \caption{Kiel diagram in [$T_{\text{eff}}$,$\log g$] space with a coloured residual scale (BHS - literature) for the Morton et al. (2016) catalogue. Training set coverage is highlighted by open black circles.}
     \label{fig:mortonHR}
\end{figure}

In Fig. \ref{fig:logg_comp}, for all values and in particular \cite{morton_2016} we notice a split around solar $\log g$ for our BHS predictions. Between $\log g = 4.4$ and 4.6 dex, we overestimate $\log g$ while \cite{morton_2016} underestimates it. Now if we consider the residual in [$T_{\text{eff}}$, $\log g$] space shown in Fig. \ref{fig:mortonHR}, we have smaller residuals in that region. In this case it seems to be a joint effect. Our models underestimate mass and radius while \cite{morton_2016} seems to overestimate them leading to a negative residual. This aligns with the opposite shapes seen in Fig. \ref{fig:logg_comp}. An underestimate in both radius and mass of around the same scale would lead to an overestimate in $\log g$ due to its stronger inverse dependence with $R^2$ and vice versa. However, this bias is also visible in our BHS $\log g$'s for non-Morton values while the literature does not present it. Thus, we can conclude that our models slightly underestimate mass and radius for sub-solar dwarfs. This is most likely due to a lack of training points in this region as shown in Fig. \ref{fig:mortonHR}. 

The underestimation of mass and radius for sub-solar dwarfs arises because the model’s learnt mapping flattens in regions with few training examples.
In multidimensional $T_{\text{eff}}-\log g - [\text{Fe}/\text{H}]$ space, the regression surface pulls predictions towards the dense solar-like ridge, which, for cool high-gravity stars, lies above them in $T_{\text{eff}}$ and below them in $\log g$.
That projection leads to smaller inferred radii and masses than their reference values. We have an asymmetric information content: cooler dwarfs contribute less to the training objective due to their scarcity and their higher uncertainties in input parameter space.

This behaviour is consistent with expectations from model regularisation and the under-representation of cool dwarfs in asteroseismic and eclipsing-binary samples that make up our training set, where observational selection favours larger, more luminous stars.

A similar data-scarcity bias may also influence the opposite end of our training range, flattening the model’s learnt relation above approximately 1.4 $M_\odot$ and 2.0 $M_\odot$. This would again lead to underestimates at the high-mass and large-radius end. Figs. \ref{fig:zoomed_res_ML} and \ref{fig:full_res} may hint at such behaviour for radius, although this remains less conclusive due to the smaller number of stars in that regime.

These findings motivate future work aimed at improving the coverage of our training database. Aligning its distribution more closely with the underlying stellar population will likely be key to achieving both higher agreement and better-calibrated uncertainty estimates.

In the region around $T_{\text{eff}} \approx 6000-7000$ K and $\log g \approx 4.2-4.4$ dex, corresponding to late-F to early-G turn-off stars, our BHS model predicts slightly larger masses and radii than the values reported by \cite{morton_2016} as seen in Fig. \ref{fig:mortonHR}. This produces a positive residual (BHS – literature) and a corresponding slight underestimate in our inferred $\log g$.

This behaviour is consistent with expectations from isochrone-based stellar inference. Near the main-sequence turn-off, evolutionary tracks become densely packed and partially degenerate, such that inferred masses and radii can be sensitive to priors and interpolation choices. In these regimes, Bayesian frameworks tend to favour unevolved solutions when observational constraints are insufficient to robustly distinguish evolutionary states. The systematic offset observed relative to \citet{morton_2016} is therefore plausibly attributable to methodological differences rather than deficiencies in either dataset.

The absence of this bias in non-Morton stars supports the interpretation that the offset originates from the isochrone method rather than from our model. The apparent agreement in Morton's $\log g$ seen in Fig. \ref{fig:logg_comp} is likely due to imposed self-consistency with its own inputs. 

\subsubsection{Physical consistency of the predictions}

Another result that points towards improved mass and radius predictions, at least within the accuracy of the input parameters, is the enhanced theoretical consistency of our models. Although mass and radius are predicted independently, they yield a lower overall scatter in $\Delta \log g$ than the literature values. For our BHS predictions, we find $\sigma_{\Delta \log g}=0.012$ dex, compared to $0.038$ dex for the literature. This tighter consistency indicates that our predicted masses and radii obey the expected physical relation more closely, suggesting that the model captures coherent stellar structure trends even without explicitly enforcing them.

Overall, our models achieve excellent agreement with literature values while maintaining stronger internal consistency, reinforcing their reliability for characterising host stars across a broad range of stellar parameters.

\section{Conclusion}

We have presented a probabilistic, data-driven framework for estimating stellar mass and radius from commonly available spectroscopic and photometric observables. Bayesian hierarchical stacking combines models with complementary statistical assumptions while propagating measurement uncertainties and accounting for correlations among the inputs. The resulting method is intended as a complementary, catalogue-scale approach to grid-based stellar modelling: it learns the relations represented by a curated reference sample and can be applied rapidly without selecting stellar-evolution tracks for each target. 

On the held-out test set, BHS obtains mean absolute relative differences of 3.5 per cent in mass and 2.1 per cent in radius. These values are lower than those obtained with the empirical relations and the previous non-Bayesian machine-learning approach evaluated on the same sample. The probabilistic metrics also indicate competitive predictive performance, although the coverage and normalised residuals show that the resulting predictive intervals are generally conservative. 

For 44 Kepler LEGACY stars outside the model-development sample, the BHS masses differ from the BASTA estimates by a MARD of 4.8 per cent, compared with 8.5 per cent between the empirical relations and BASTA. This closer agreement does not establish which method is nearer to the unknown physical masses, but supports the use of BHS and grid-based inference as complementary approaches with different assumptions and input requirements. 

When applied to approximately 2400 exoplanet host stars, the BHS predictions differ from the adopted literature values by a MARD of 2.7 per cent for both mass and radius. The independently inferred masses and radii also produce three times less scatter in $\Delta\log g$ than the corresponding literature values. This result indicates greater internal consistency. By providing catalogue-scale estimates with propagated uncertainties, the framework can support the characterisation of large stellar and exoplanet samples as new spectroscopic surveys become available. 

The present training sample is dominated by F- and G-type stars. Future work will focus on expanding the training database towards M dwarfs and higher-mass stars. We also plan to include this work as part of a larger tool that also predicts stellar age.

\section*{Data availability}
The code is publicly available at \url{https://github.com/lamirelm/BayeStarML}, and an online implementation is planned within the Spanish Virtual Observatory, extending the existing empirical-relation tool available at \url{http://sdc.cab.inta-csic.es/empiricalRelationsMR/}.

\begin{acknowledgements}
      A.M. acknowledges funding support from grant PID2023-149439NB-C44 funded by MCIN/AEI/10.13039/501100011033/ and from FEDER "Una manera de hacer Europa, EU", and from Generalitat Valenciana in the frame of the GenT Project ESGENT-CIESGT2025. We would also like to thank the referee and editor for their valuable input. 
\end{acknowledgements}

\bibliographystyle{bibtex/aa}
\bibliography{ref}

\begin{appendix}
\nolinenumbers
\section{Additional model specification}

\subsection{Bayesian hierarchical stacking}
\label{app:bhs}

The stacking weights are estimated by maximising the leave--one--out (LOO) expected log predictive density, approximated using Pareto--smoothed importance sampling (PSIS--LOO). In practice, we provide as input the matrix of pointwise PSIS--LOO log predictive densities, $\mathbf{lpd}$, whose elements $\mathrm{lpd}_{ik}$ are indexed by observation $i$ and model $k$. The optimisation problem can be written as
\begin{equation}
\widehat{\mathbf{w}}
=
\arg\max_{\mathbf{w}}
\sum_{i=1}^{n}
\log\!\left(
\sum_{k=1}^{K}
w_k(\mathbf{x}_i)\,
\exp\{\mathrm{lpd}_{ik}\}
\right),
\label{eq:loo_objective}
\end{equation}
where $n$ is the number of observations. Here, $\mathrm{lpd}_{ik}$ corresponds to the LOO log predictive density of model $k$ for observation $i$, i.e. $\mathrm{lpd}_{ik}=\log \hat p_k(y_i \mid \mathbf{x}_i,\mathcal{D}_{-i})$, with $\mathcal{D}_{-i}$ denoting the training dataset with the $i$-th observation removed \citep{loo,bhs}.

Equation~\ref{eq:loo_objective} expresses the conventional stacking optimisation problem. In our Bayesian implementation, the quantity inside this objective is instead included as a log potential, and the posterior distribution of the feature-dependent weight coefficients is sampled under the hierarchical priors defined below.

All input parameters are first standardised using the training-set means and standard deviations. For the stacking model, each standardised predictor is additionally decomposed into its negative and positive deviations from the training-set median,
\begin{equation}
\boldsymbol{\phi}(\mathbf{x}_i)
=
\left[
\min(\mathbf{x}_i-\widetilde{\mathbf{x}},0),
\max(\mathbf{x}_i-\widetilde{\mathbf{x}},0)
\right].
\end{equation}
This doubles the feature dimension and allows the weights to respond differently to values below and above the median. The same training-set transformation is applied to all evaluation samples.

Unlike standard stacking with fixed global weights, hierarchical stacking allows the weights to vary across the predictor space. In our implementation, this is achieved by introducing latent, unconstrained variables (logits) that encode the relative preference for each model as a function of the input features. The logits are modelled as linear functions of the feature vector, allowing the stacking weights to vary smoothly across the predictor space, and are subsequently mapped to the probability simplex using a softmax transformation, ensuring positivity and unit normalisation. Concretely, for observation $\mathbf{x}_i$ we define
\begin{equation}
\eta_k(\mathbf{x}_i)
=
\boldsymbol{\phi}(\mathbf{x}_i)^\top\boldsymbol{\beta}_k,
\qquad k=1,\ldots,K-1,
\label{eq:linear_logits}
\end{equation}
and set $\eta_K(\mathbf{x}_i)=0$ as a reference to ensure identifiability. The corresponding feature-dependent stacking weights are then
\begin{equation}
w_k(\mathbf{x}_i)
=
\frac{\exp\{\eta_k(\mathbf{x}_i)\}}
{\sum_{\ell=1}^{K} \exp\{\eta_\ell(\mathbf{x}_i)\}}.
\label{eq:softmax}
\end{equation}

To regularise the feature dependence of the weights, we use a hierarchical (partial pooling) prior on the regression coefficients. Specifically, each coefficient vector $\boldsymbol{\beta}_k$ is modelled as
\begin{equation}
\beta_{k j} \sim \mathcal{N}(\mu_k,\sigma_k^2),
\qquad j=1,\ldots,d,
\label{eq:beta_prior}
\end{equation}
where $d$ is the number of continuous features, and the model-specific parameters $(\mu_k,\sigma_k)$ are assigned common-form priors,
\begin{equation}
\mu_k\sim\mathcal{N}(0,1),
\qquad
\sigma_k\sim\mathrm{HalfNormal}(0.5).
\label{eq:hyperpriors}
\end{equation}

This construction partially pools the feature coefficients within each component model towards a model-specific mean, reducing overfitting while allowing the weights to vary across the predictor space.

\subsection{BART}
\label{app:bart}

A complete BART model can be expressed as
\begin{equation}
    y \sim \mathcal{N}\!\left( \sum_{j=1}^{m} g_j(\mathbf{X}; T_j, M_j),\, \sigma^2 \right),
    \label{eq:BART}
\end{equation}
where $T_j$ denotes the tree structure (including splitting variables and cut-points) and $M_j$ represents the set of terminal-node values for tree $j$ \citep{quiroga2022bart}. 

Following standard BART practice, independent priors are placed on the tree structures and leaf parameters. In particular, we adopt the depth prior
\begin{equation}
    \Pr(\text{non-terminal at depth } d) = \alpha (1 + d)^{-\beta},
    \qquad \alpha \in (0,1), \; \beta \ge 0,
    \label{eq:depth}
\end{equation}
which favours shallow trees and acts as a regularisation mechanism \citep{bart}. The remaining priors on splitting variables, cut-points, and terminal-node values follow the default BART specification as implemented in the PyMC framework \citep{quiroga2022bart}.

The number of trees $m$ controls the overall model complexity, with predictive performance typically improving as $m$ increases before reaching a plateau. In our implementation, we find that $m = 250$ provides an effective balance between predictive performance and computational efficiency.

\subsection{TP-HSGP}
\label{app:gp}

To prevent pathological covariance structures and to regularise the length-scales, we set adaptive priors on the kernel hyperparameters.  
For each input dimension, the mean and standard deviation of an inverse-gamma prior on the length-scale \(\ell_d\) are determined from the distribution of pairwise distances between training points in that dimension.  
This ensures that \(\ell_d\) is constrained to vary within a range appropriate to the spacing of the training data—neither much shorter than the smallest inter-point separation nor longer than the overall span of the data. The kernel amplitude parameters \(\eta_d\) follow weakly informative gamma priors (\(\alpha = 2,\, \beta = 1\)), providing additional regularisation of the covariance amplitude.  
These adaptive priors stabilise inference and prevent overfitting by discouraging degenerate or near-singular covariance matrices.

In our implementation, we perform full Bayesian inference over the kernel hyperparameters and inducing variables rather than relying on point estimates.  
We place the adaptive priors described above and sample from the joint posterior
\begin{equation}
p(\theta, \mathbf{u} \mid \mathcal{D}) \;\propto\; p(\mathbf{y}\mid \mathbf{X}, \theta, \mathbf{u})\, p(\mathbf{u}\mid \theta)\, p(\theta),
\label{eq:joint_posterior}
\end{equation}
where $\theta=\{\eta_d,\ell_d\}_{d=1}^{4}$ denotes the kernel hyperparameters and \(\mathbf{u}\sim \mathcal{N}(\mathbf{0}, K_{uu})\) are the inducing variables \citep[e.g.][]{titsias2009variational, snelson2006sparse}.  
Predictive distributions marginalise over \(\theta\) and \(\mathbf{u}\) by averaging the GP conditionals across posterior samples, thereby propagating uncertainty in both the covariance function and the inducing representation to the final estimates.  
For completeness, hyperparameters in standard GP models are often obtained by maximising the log marginal likelihood \citep[see][]{rasmussen2006gpml}, but full Bayesian inference provides a more principled treatment of uncertainty in this context.
\onecolumn

\subsection{HBNN}
\label{app:hbnn}

The first level focuses on the covariance between the input parameters. To do so, we model the latent ‘true’ stellar parameters $\mathbf{x_{latent}}$ as a multivariate normal distribution with a LKJ Cholesky \citep{LKJ_2009} prior on its covariance matrix. $\mathbf{x_{latent}}$ is inferred from the observed parameters represented alongside their errors as normal likelihoods. This allows us to capture the correlation between the input parameters. 

For the latent inputs we place a zero-mean multivariate normal prior with unknown covariance \(\boldsymbol{\Sigma}\):

\begin{equation}
\mathbf{x_{\text{latent}}}^\star \sim \mathcal{N}\!\big(\mathbf{0},\ \boldsymbol{\Sigma}\big), \qquad 
\boldsymbol{\Sigma} = \mathbf{D}\,\mathbf{R}\,\mathbf{D},
\label{eq:covar}
\end{equation}

where \(\mathbf{R}\) has an LKJ prior on the correlation matrix (shape parameter \(\eta=2\)) and the marginal scales in \(\mathbf{D}=\mathrm{diag}(\sigma_1,\ldots,\sigma_4)\) have half-normal priors.

Then the observation model links measurements \(\mathbf{x}\) and their reported \(1\sigma\) errors \(\mathbf{s}\) to the latent quantities:
\begin{equation}
\mathbf{x} \mid \mathbf{x_{\text{latent}}} \sim \mathcal{N}\!\big(\mathbf{x_{\text{latent}}},\ \mathrm{diag}(\mathbf{s}^2)\big).
\label{eq:latent}
\end{equation}

The latent inputs are passed through two hidden layers with leaky-ReLU activations of fixed slope $\alpha=0.01$. We assign independent Gaussian priors $N(0,1)$ to the weights and biases.

\section{Model metrics table}
\label{app:metrics}

\begin{table}[h]
\caption{Independent test-set performance for (a) stellar mass and
(b) stellar radius.}
\label{tab:combined_perf}
\centering
\setlength{\tabcolsep}{3pt}
\begin{subtable}[t]{0.575\textwidth}
\centering
\resizebox{\linewidth}{!}{%
\begin{tabular}{lcc cccc}
\toprule
& \multicolumn{2}{c}{Point} & \multicolumn{4}{c}{Probabilistic} \\
\cmidrule(lr){2-3}\cmidrule(lr){4-7}
Model & MARD (\%) & MRD (\%) & LPD $\uparrow$ &
CRPS $\downarrow$ ($M_\odot$) & 68\% cov. & $\sigma_z$ \\
\midrule
TP-HSGP & 3.7 & 0.2 & $1.21\pm0.06$ & $0.037\pm0.003$ & $86\pm4$ & 0.76 \\
HBNN & 3.5 & 0.4 & $1.20\pm0.11$ & $0.035\pm0.003$ & $84\pm4$ & 0.98 \\
BART & 3.9 & -0.7 & $1.15\pm0.04$ & $0.039\pm0.002$ & $88\pm3$ & 0.62 \\
BHS & 3.5 & -0.05 & $1.25\pm0.06$ & $0.035\pm0.003$ & $85\pm4$ & 0.79 \\
ER {\tiny\citep{moya1}} & 6.3 & 1.1 & $0.1\pm0.3$ & $0.058\pm0.005$ & $46\pm5$ & 1.83 \\
AI {\tiny\citep{moya2}} & 3.7 & 0.1 & \multicolumn{4}{c}{---} \\
\bottomrule
\end{tabular}%
}
\caption{Mass predictions.}
\label{tab:mass_perf}
\end{subtable}\hfill
\begin{subtable}[t]{0.41\textwidth}
\centering
\resizebox{\linewidth}{!}{%
\begin{tabular}{cc cccc}
\toprule
\multicolumn{2}{c}{Point} & \multicolumn{4}{c}{Probabilistic} \\
\cmidrule(lr){1-2}\cmidrule(lr){3-6}
MARD (\%) & MRD (\%) & LPD $\uparrow$ &
CRPS $\downarrow$ ($R_\odot$) & 68\% cov. & $\sigma_z$ \\
\midrule
2.1 & 0.001 & $1.52\pm0.06$ & $0.028\pm0.002$ & $88\pm3$ & 0.72 \\
2.2 & -0.3 & $1.52\pm0.10$ & $0.029\pm0.002$ & $69\pm5$ & 1.04 \\
2.3 & -0.1 & $0.87\pm0.01$ & $0.043\pm0.001$ & $99\pm1$ & 0.32 \\
2.1 & 0.05 & $1.51\pm0.06$ & $0.028\pm0.002$ & $88\pm3$ & 0.69 \\
2.7 & -0.3 & $1.38\pm0.09$ & $0.034\pm0.003$ & $78\pm4$ & 0.9 \\
2.7 & 0.5 & \multicolumn{4}{c}{---} \\
\bottomrule
\end{tabular}%
}
\caption{Radius predictions.}
\label{tab:radius_perf}
\end{subtable}
\tablefoot{Metric definitions are given in Sect.~\ref{sec:eval_met}. Higher LPD and lower CRPS indicate better performance. Uncertainties are standard errors of the mean for LPD and CRPS, and binomial standard errors for the empirical 68 per cent coverage. The AI model of \citet{moya2} provides no predictive uncertainties; probabilistic metrics are therefore unavailable. Radius results follow the model order shown in panel (a).}
\end{table}

\section{Test set performance plot}
\label{app:test_set_fig}

\begin{figure}[ht]
\centering
\begin{subfigure}[b]{0.48\textwidth}
\centering
\includegraphics[width=0.7\linewidth,height=0.2\textheight,keepaspectratio]{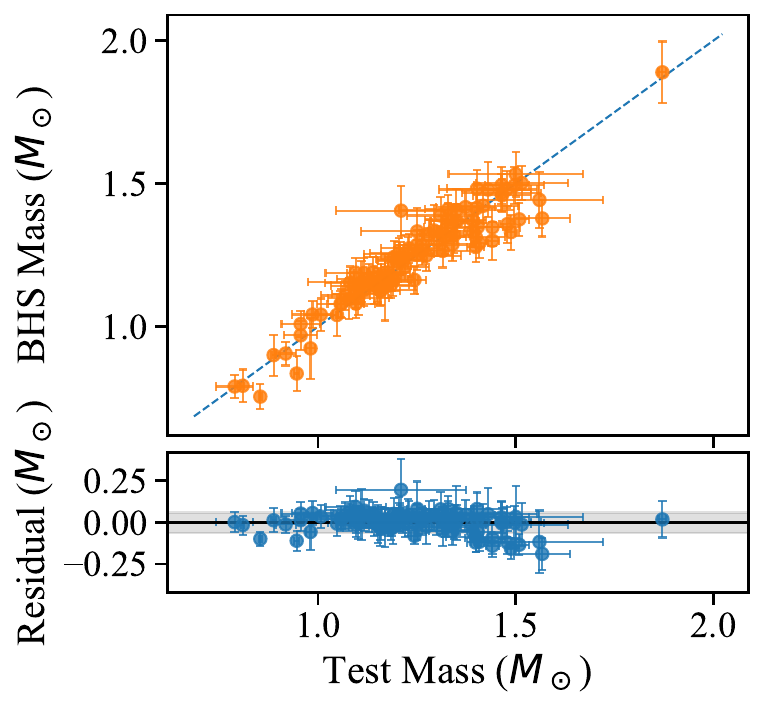}
\caption{Mass}
\label{fig:test_mass}
\end{subfigure}\hfill
\begin{subfigure}[b]{0.48\textwidth}
\centering
\includegraphics[width=0.7\linewidth,height=0.2\textheight,keepaspectratio]{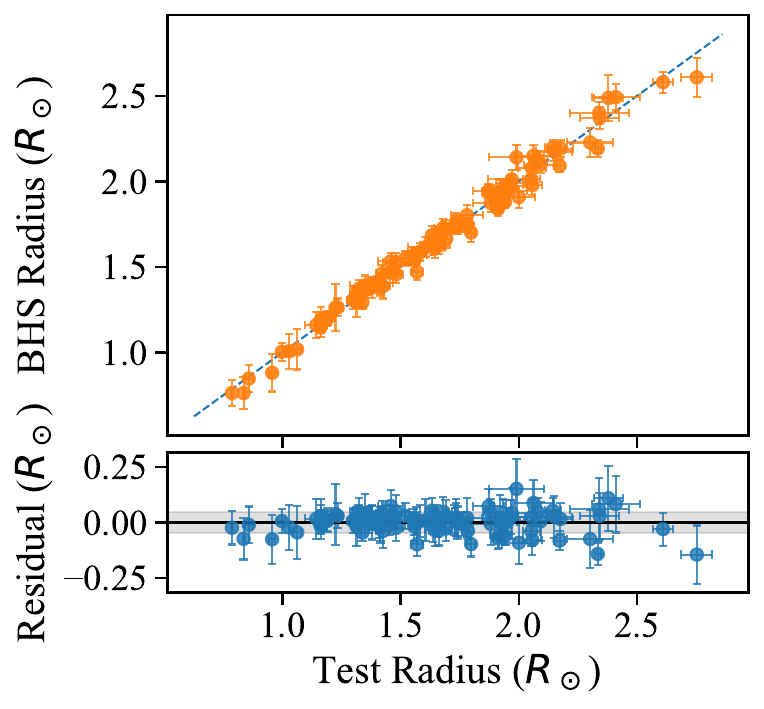}
\caption{Radius}
\label{fig:test_radius}
\end{subfigure}
\caption{BHS predictions against the adopted test-set reference values (orange) and the corresponding residuals, defined as prediction minus reference value (blue).}
\label{fig:test}
\end{figure}

\end{appendix}

\end{document}